\documentclass[longauth]{aa} 

\usepackage{graphicx}
\usepackage{color}
\usepackage{amsmath}    
\usepackage{amssymb}    
\usepackage{natbib}
\bibpunct{(}{)}{;}{a}{}{,} 
\usepackage{longtable}
\usepackage{lscape}
\usepackage{txfonts}
\usepackage{morefloats}
\usepackage[normalem]{ulem}	
\newcommand\mr[1]{{{\color{mmr} #1}}}

\definecolor{mmr}{rgb}{0.7,0.0,0.2}

\definecolor{mmg}{rgb}{0.0,0.7,0.3}

\begin{document}
\title{International observational campaign of the 2014 eclipse of EE\, Cep
\thanks{Tables \ref{TA1} -- \ref{TA29} are available at the CDS via anonymous
ftp to cdsarc.u-strasbg.fr \mr{(XXX.XX.XXX.X)} or via
http://cdsweb.u-strasbg.fr/cgi-bin/qcat?J/A+A/\mr{XXX/XXX}}}

\author{D. Pie\'nkowski\inst{1}
\and C. Ga{\l}an\inst{1}
\and T. Tomov\inst{2}
\and K. Gazeas\inst{3}
\and P. Wychudzki\inst{2}
\and M. Miko{\l}ajewski\inst{2}		
\and D. Kubicki\inst{2}
\and B. Staels\inst{4}		
\and S. Zo{\l}a\inst{5,6}
\and P. Pako\'nska\inst{5}
\and B. D{\c e}bski\inst{5}
\and T. Kundera\inst{5}
\and W. Og{\l}oza\inst{6}
\and M. Dr\'o\.zd\.z\inst{6}
\and A. Baran\inst{6}
\and M. Winiarski\inst{6}
\and M. Siwak\inst{6}
\and D. Dimitrov\inst{7}
\and D. Kjurkchieva\inst{8}
\and D. Marchev\inst{8}
\and A. Armi\'nski\inst{9}
\and I. Miller\inst{10}
\and Z. Ko{\l}aczkowski\inst{11}
\and D. Mo\'zdzierski\inst{11}		
\and E. Zahajkiewicz\inst{11}
\and P. Bru\'s\inst{11}
\and A. Pigulski\inst{11}		
\and T. Smela\inst{9}		
\and E. Conseil\inst{12}
\and D. Boyd\inst{13}
\and G. J. Conidis\inst{14}		
\and I. Plauchu-Frayn\inst{15}
\and T. A. Heras\inst{16}
\and E. Kardasis\inst{17}
\and M. Biskupski\inst{9}		
\and R. Kneip\inst{30}		
\and L. Hamb\'alek\inst{18}
\and T. Pribulla\inst{18}
\and E. Kundra\inst{18}
\and J. Nedoro\v{s}\v{c}ik\inst{18}
\and J. Lopatovsk\'y\inst{18}
\and Z. Garai\inst{18}
\and D. Rodriguez\inst{19}
\and T. Kami\'nski\inst{20}
\and F. Dubois\inst{21}	
\and L. Logie\inst{21}
\and A. Capetillo Blanco\inst{30}
\and P. Kankiewicz\inst{22}
\and E. \'Swierczy\'nski\inst{2}
\and M. Martignoni\inst{23}		
\and I. Sergey\inst{24}
\and J. Kare Trandem Qvam\inst{25}
\and E. Semkov\inst{7}
\and S. Ibryamov\inst{7,8}
\and S. Peneva\inst{7}
\and J. -L. Gonzalez Carballo\inst{26}
\and J. Ribeiro\inst{27}	
\and S. Dean\inst{28}
\and G. Apostolovska\inst{29}
\and Z. Donchev\inst{7}
\and L. Corp\inst{30,31}
\and P. McDonald\inst{30}
\and M. Rodriguez\inst{32}
\and A. Sanchez\inst{33}		
\and K. Wiersema\inst{34}
\and E. Conseil\inst{30}
\and J. Menke\inst{35}
\and I. Sergey\inst{36}
\and N. Richardson\inst{37}
	}


\institute{		
Nicolaus Copernicus Astronomical Centre PAS, Warsaw, Poland\\
\email{dapien@camk.edu.pl, cgalan@camk.edu.pl}
\and 
Nicolaus Copernicus University, Toru\'n, Poland
\and 
Section of Astrophysics, Astronomy and Mechanics, Department of Physics, National and Kapodistrian University of Athens, GR 15784 Zografos, Athens, Greece
\and 
Sonoita Research Observatory/AAVSO, USA
\and 
Astronomical Observatory, Jagiellonian University, Cracow, Poland
\and 
Mount Suhora Astronomical Observatory, Cracov Pedagogical University, ul. Podchorazych 2, PL-30-084 Cracow, Poland
\and 
Institute of Astronomy and National Astronomical Observatory, Bulgarian Academy of Sciences, Sofia, Bulgaria
\and 
Department of Physics, Shumen University, Shumen, Bulgaria
\and 
PTMA, Szczecin, Poland
\and 
Furzehill House, Ilston, Swansea, SA2 7LE, UK
\and 
Astronomical Institute of the Wroc{\l}aw Univeristy, Wroc{\l}aw, Poland
\and 
Club d'astronomie de Mont-Bernenchon, France
\and 
BAA Variable Star Section, West Challow Observatory, Oxfordshire, UK
\and 
Department of Physics and Astronomy, York University, Toronto, Ontario, Canada
\and 
Instituto de Astronomıa, Universidad Nacional Autonoma de Mexico, Ensenada, B.C., Mexico
\and 
Observatorio Las Pegueras, NAVAS DE ORO (Segovia), Spain
\and 
Hellenic Amateur Astronomy Association, Athens, Greece
\and 
Astronomical Institute, Slovak Academy of Sciences, Tatransk\'a Lomnica, Slovak Republic
\and 
Guadarrama Observatory, MPC458
\and 
Center for Astrophysics, Harvard \& Smithsonian, 60 Garden Street, Cambridge, MA, USA
\and 
Astrolab Iris, Zillebeke Belgium
\and 
Institute of Physics, Astrophysics Division, Jan Kochanowski University, Kielce, Poland
\and 
Stazione Astronomica Betelgeuse, Magnago, Milano, Italy
\and 
Private Observatory, Maladziechna, Belarus
\and 
The Faculty of Mathematics and Natural Sciences, Department of Physics, University of Oslo, Norway
\and 
Cerro del Viento Observatory, Badajoz, Spain (MPC I84)
\and 
Observatorio do Instituto Geografico do Exercito - Lisboa, Portugal
\and 
AstroCamp Observatory, Nerpio, Spain (MPC I89)
\and 
Institute of Physics, Faculty of Science, Ss. Cyril and Methodius University, Skopje, FYR of Macedonia
\and 
AAVSO, Cambridge, MA 02138, USA
\and 
Groupe Europ\'een d’Observations Stellaires (GEOS), Bailleau l’Ev\^eque, France
\and 
Private Observatory Madrid-Ventilla, Spain, MPC J30
\and 
Gualba Observatory, Spain
\and 
Department of Physics and Astronomy, University of Leicester, UK
\and 
Menke Scientific, Barnesville, MD 20838, USA
\and 
Amateur society "Astroblocknote", Minsk, Belarus
\and 
Mont M\'egantic Observator, University of Montreal, Canada
}

   \date{Received xxxxx xx, xxxx; accepted xxxxx xx, xxxx}

  \abstract
   {EE Cep is one of few eclipsing binary systems with a dark, dusty disk around an invisible object similar to $\varepsilon$\,Aur.  The system is characterized by grey and asymmetric eclipses every 5.6 yr, with a significant variation in their photometric depth, ranging from $\sim 0^m.5$ to $\sim 2^m.0$.}
   {The main aim of the observational campaign of the EE Cep eclipse in 2014 was to test the model of disk precession \citep{Gal2012}. We expected that this eclipse would be one of the deepest with a depth of $\sim 2^m.0$.  }
   {We collected multicolor observations from almost 30 instruments located in Europe and North America. This photometric data covers 243 nights during and around the eclipse. We also analyse the low- and high-resolution spectra from several instruments.}
   {The eclipse was shallow with a depth of $0^m.71$ in $V$-band. The multicolor photometry illustrates small color changes during the eclipse with a total amplitude of order $\sim+0$\fm$15$ in $B - I$ color index. The linear ephemeris for this system is updated by including new times of minima, measured from the three most recent eclipses at epochs $E$\,=\,9,\,10 and 11. New spectroscopic observations were acquired, covering orbital phases around the eclipse, which were not observed in the past and increased the data sample, filling some gaps and giving a better insight into the evolution of the $H_{\alpha}$ and \ion{Na}{I} spectral line profiles during the primary eclipse.}
   {The eclipse of EE Cep in 2014 was shallower than expected $0^m.71$ instead of $\sim2$\fm$0$. This means that our model of disk precession needs revision.}

   \authorrunning {D. Pie\'nkowski et al.}

   \titlerunning {International observational campaign of the 2014 eclipse of EE\, Cep}


   \maketitle
%


\section{Introduction}
The 11-th magnitude system EE\,Cep (BD$+55 2693$) is a member of a rare class of
binary systems, in which the eclipses are caused by a dark, dusty disk
surrounding the orbiting companion.  The precursor of this group is the
extremely long-period (27.1 yr) eclipsing binary system $\varepsilon$\,Aur
\citep[see][]{Gui2002}.  For a long time, these two systems were the only
ones in this class of systems.  Nowadays, more than a dozen such systems
with similar properties are known, since many researchers have been reported
(see eg.  \citet{Lip2016},
\citet{Gar2016}, \citet{Rat2015}, \citet{KeMa2015}, \citet{Sco2014} and
references in \citet{Gal2014}). Among these systems, there is a large
diversity in orbital period duration -- from days to decades.  Most of these
systems have weak observational records, frequently with one or
a few eclipses observed so far. The interpretation of their 
nature is therefore often uncertain and requires verification through further
observations and analysis. An example is the case of M2-29, which we
ascribed into this group \citep{Gal2014}, based on \citet{Haj2008} analysis
of a single eclipse observed by OGLE and MACHO surveys.  However
\citet{Misz2011} prove that the scenario of eclipses by dusty disk is
unlikely and propose another explanation of the observed eclipse as due to condensed
dust and its evaporation around the R\,CrB star.

The EE\,Cep system stands out in this group with a relatively long orbital
period (5.6\,yr) after $\varepsilon$\,Aur and TYC
2505-672-1 (69.1\,yr, \citet{Lip2016}).  It has a well documented long
history of eclipses showing very unusual behaviour, for over 13 orbital
epochs. The research on this object dates back to the middle
of the last century.  It was discovered as a variable in 1952 (epoch
$E$\,=\,0) by \citet{Rom1956}. confirmed soon after by \citet{Web1956},
who reported the singular observation of the change in its brightness during
the previous eclipse in 1947 ($E$\,=\,$-1$).  The eclipsing nature of the light curve
 was established with photometric observations of the
subsequent events in 1958, 1964 and 1969 \citep{Mei1973}. Henceforth all
consecutive, primary eclipses were observed, but no traces of the secondary
eclipse have ever been recorded.

The first analytical model of the system was proposed by \citet{Mei1975}.  
He assumed that the B-type primary star is eclipsed by an M-type red
giant. Pulsations of the giant should cause changes in depth of the
eclipses whereas its inflated atmosphere could account for atmospheric
wings.  This model was questioned when the first multi-color observations
using the $R$ and $I$ pass-bands of Johnson's photometric system were
obtained in Piwnice Observatory in 1997 \citep{MiGr1999}.  Very small color
changes, observed during the eclipse, against the interpretation
that the eclipsing component is a red giant.  \citet{MiGr1999} proposed
another model in which the eclipses could be caused by an invisible,
cold object -- a dark disk around a low-luminosity central star or a close binary.
From an analysis of the color indices, they constrained the basic parameters of
the primary component: a hot star of spectral type B5, effective temperature
14300\,K, radius 10\,$R_{\odot}$, and distance of 2.75\,kpc. Gaia DR2
\citep[][2018]{Gai2016} brought an independent parallax $\pi = (0.503
\pm 0.032) \times 10^{-3}$\,arcsec,
indicating that EE Cep system is closer, at $1.99\pm0.13$\,kpc.


\begin{figure*}
\centering
  \includegraphics[width=17cm]{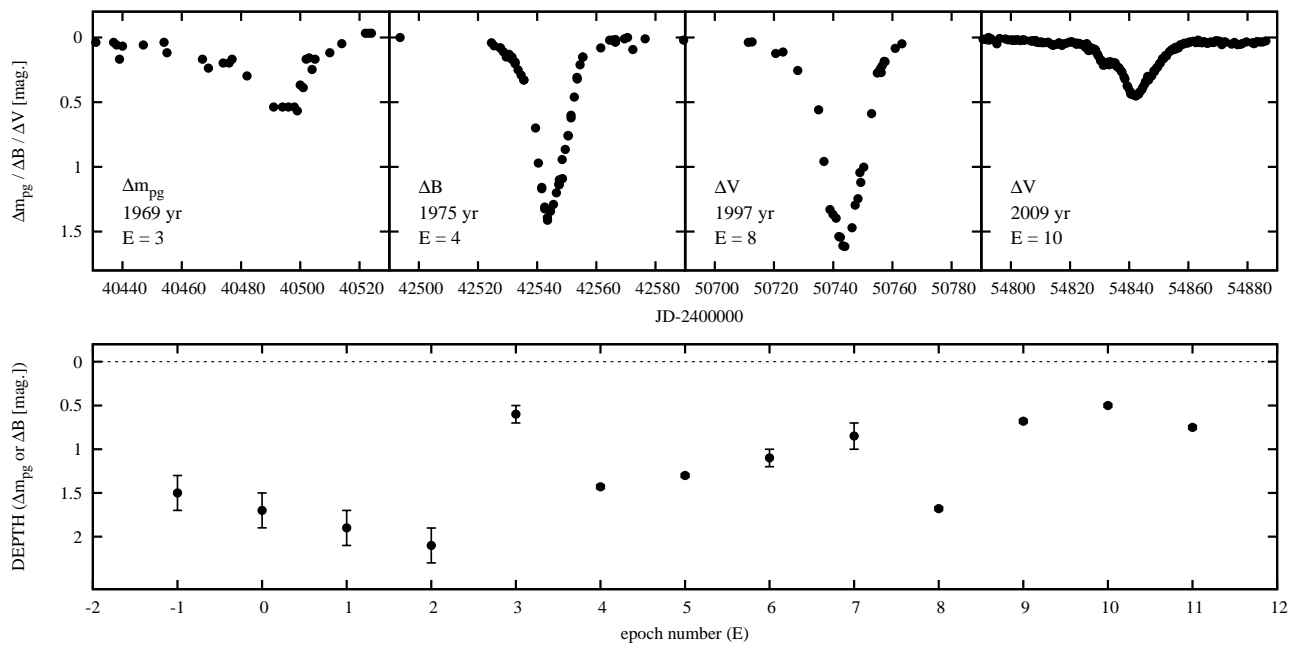}
\caption{{\sl Top}: Four representative light curves of the eclipses in
EE\,Cep selected among the best quality so far light curves with good time
coverage -- two deep ($E$\,=\,4\,and\,8) and two shallowest
($E$\,=\,3\,and\,10).  {\sl Bottom}: Time dependence relation for the depth
of eclipses.}
  \label{eclipses}
\end{figure*}

The most striking feature of the EE\,Cep minima is the significant variation
in their depth, from $\sim 0\fm5$ to $\sim 2\fm0$ (Fig.\ref{eclipses}).  
The top panel in Fig. \ref{eclipses} shows
the representative light curves of four minima: typical examples of deep
minima observed in 1975 ($E$\,=\,4) and 1997 ($E$\,=\,8), the shallowest (so
far) eclipse observed in detail during 2008/9 ($E$\,=\,10), and the
exceptional eclipse with flat bottom during its central part in 1969
($E$\,=\,3).  The bottom panel of the same figure shows the variation in time of the eclipses
The changes in the depth are
accompanied by variations of the total duration of the eclipses \citep{Gra2003}.
All eclipses are characterized by a similar asymmetry -- the descending
branches are longer than ascending ones.  It is possible to distinguish in
light curves repeatable phases during the eclipses: more or less
sloped-bottomed transit during the central part of the eclipse is preceded
and followed by the real ingress and egress, and atmospheric wings caused by
semitransparent external parts of the eclipsing body.  To explain this
unusual behavior, \citet{MiGr1999} suggested that the eclipses are caused by
the disk which is opaque in its interior and semitransparent in outer regions and
which has a varying inclination to the line of sight during different
eclipses due to precession.  In this scenario, the unique flat-bottomed
eclipse, observed in 1969 ($E$\,=\,3),
can be explained by a nearly edge-on and non-tilted projection of the disk.

To test this model we needed good temporal coverage and high quality
photometric and spectroscopic observations. Obtaining dense enough sampling
of the eclipses would be possible only through the collective effort of a
large number of observers in various locations around the world.  For the
events in 2003 and 2008/9, we organized extensive, international observing
campaigns (\citet{Mik2003}; \citet{Gal2008}), which attracted the attention
of several dozens of observers.  Analysis of the high-resolution spectra of
these campaigns \citep{Gal2012} confirmed that the primary component is a
B5III star rotating at very high velocity ($\sim$350 km/s) which causes
inhomogeneous temperature and flux distribution on its surface due to the
\citet{Zei1924} effect. Recent photometric results complemented by the historical
light curves \citep{Gra2003} were used to calculate a model of the system
that would be able to reproduce the changing shapes of the eclipses caused
by the disk precession and taking into account a mechanism which would explain the unusually different eclipses at epochs
$E$\,=\,3 and 8 (see Fig.\ref{eclipses} -- bottom).  The the best solution obtained for the disk precession was $P_{prec} \sim11 - 12 P_{orb}$
\citep{Gal2012}.  According to this model, the eclipse in 2014 (E =11)
should attain a large, close to $2\fm0$.  This model was based on
observations obtained in a time interval almost exactly equal to the
derived precession period. Additional photometric and spectroscopic
observations were needed for its verification.

In this paper, we present photometric and spectroscopic data
obtained with the new campaign that constitutes a basis for future
papers with quantitative analysis.  Section\,2 describes technical details
of the campaign in 2014, including a compilation of instruments
involved in the observations and a
description of techniques used to combine data from various photometric
systems.  Section\,3 presents the results of the campaign, i.e.  the
light curves and spectra.  A brief discussion of the conclusions
from obtained results is made in Section\,4.

\section{Observations}

Our call for observations \citep{Gal2014} gained a wide response from
observers, using 30 telescopes located in Europe and two in North America. 
Several science facilities and organizations were involved in this campaign,
as well as a large number of amateur astronomers, who were participating in
an independent campaign, coordinated by AAVSO \citep{Waa2014}.  Technical
details of telescopes are listed in Table\,\ref{telescopes}.  All telescopes
were reflectors with the range from 0.2 to 2.0 m in diameter.  More than
11000 individual photometric measurements were obtained with 27 instruments
-- most using $UBV(RI)_{\rm{C}}$ pass-bands of Johnson-Cousins photometric
system and in several cases $RI$ filters more close to Johnson's realization
of these bands.  The photometric data cover 243 days from March\,22 to
November\,20, 2014.

One photometric measurement in near-infrared $JHK_{\rm{S}}$ bands was
obtained on June\,3, 2010 (HJD = 2455350.67) with the {\sl{CAIN}} infrared camera
operating on 1.5\,m Carlos S\'anchez telescope at Teide Observatory.  
The frames were reduced on flat-fields, bad pixel corrected, background
subtracted, and finally combined with the use of
{\sl{CAINDR}}\footnote{CAINDR --
Infrared Camera Reduction Software developed by R.  Berrena and J.  Acosta -
IAC v0.5 -- July 2007.} data reduction tasks
working in the {\sl{IRAF}}\footnote{IRAF is distributed by the National
Optical Astronomy
Observatories, which are operated by the Association of Universities for
Research in Astronomy, Inc., under a
cooperative agreement with the National Science Foundation.} environment, and
the photometry was performed using the {\sl{apphot}} package.  The resulted
magnitudes are shown in Table\,\ref{TA28}.

\begin{table*}
\caption{All instruments and their involvement in photometric observations of the EE\,Cep eclipses in 2014.}
\centering
\begin{tabular}{llllllll}
\hline\hline
Observatory	    & Country 	& Telescope type	    & Diameter [m]	& Bands 	& $N_i$	& $N_n$ &  Table	\\
\hline
Astrolab Iris	& Belgium	& Newton		        & 0.68		& $V$		& 25	& 25 & \ref{TA16}	\\
Athens		    & Greece	& Cassegrain 		    & 0.4		& $BV(RI)_{\rm{C}}$& 10344 & 65 & \ref{TA1}	\\
Bia\l{}k\'o{}w	& Poland	& Cassegrain		    & 0.6		& $BV(RI)_{\rm{C}}$& 150& 38 & \ref{TA5}	\\
Cerro del Viento& Spain		& Schmidt-Cassegrain	& 0.2		& $V$		& 9	& 9 & \ref{TA23}	\\
France		    & France	& Home made		        & 0.203		& $BVR	$	& 5	& 2 & \ref{TA24}	\\
Furze Hill	    & United Kingdom& Schmidt-Cassegrain& 0.35		& $BV(RI)_{\rm{C}}$& 123& 41 & \ref{TA7}	\\
Glyfada		    & Greece	& Schmidt-Cassegrain	& 0.28		& $V	$	& 44	& 44 & \ref{TA12}	\\
Guadarrama	    & Spain		& Schmidt-Cassegrain	& 0.25		& $BVRc	$	& 27	& 41 & \ref{TA15}	\\
Gualba		    & Spain		& Schmidt-Cassegrain	& 0.356		& $V(RI)_{\rm{C}}$& 3	& 1 & \ref{TA26}	\\
Horten		    & Norway	& Ritchey-Chretien	    & 0.46		& $BVR_b$	& 11	& 4 & \ref{TA21}	\\
Jan Kochanowski	& Poland	& Schmidt-Cassegrain	& 0.35		& $UBVR	$	& 20	& 5 & \ref{TA19}	\\
Las Pegueras	    & Spain     & Schmidt-Cassegrain		        & 0.35		& $V	$	& 51	& 51 & \ref{TA10}	\\
Leicester	    & United Kingdom& Corrected Dall-Kirkham& 0.508	& $BVR_{\rm{C}}$& 3	& 1 & \ref{TA25}	\\
Madrid-Ventila	& Spain		& Refractor	        & 0.06	& $V$		& 1	& 1 & \ref{TA27}	\\
Magnago		    & Italy		& Schmidt-Cassegrain	& 0.25		& $BVIc	$	& 15	& 5 & \ref{TA20}	\\
Nerpio		    & Spain		& Corrected Dall-Kirkham& 0.431		& $BVR_{\rm{C}}$& 34	& 32 & \ref{TA14}	\\
New Mexico Skies & USA		& Corrected Dall-Kirkham& 0.508		& $BV	$	& 9	& 5 & \ref{TA22}	\\
Rozhen		    & Bulgaria	& Schmidt		        & 0.7		& $UBV(RI)_{\rm{C}}$& 50& 10 & \ref{TA11}	\\
Rozhen		    & Bulgaria	& Cassegrain		    & 0.6		& $BV(RI)_{\rm{C}}$&  23& 7 & \ref{TA17}	\\
Slovak Academy	& Slovakia	& Cassegrain		    & 0.6		& $UBV(RI)_{\rm{C}}$& 139& 30 & \ref{TA6}	\\
 Sonoita	    & USA	&  Folded Newtonian	&  0.5	& $BV(RI)_{\rm{C}}$& 349& 90 & \ref{TA2}	\\
Suhora		    & Poland	& Cassegrain		    & 0.6		& $UBVRI$	& 307	& 57 & \ref{TA3}	\\
Szczecin	    & Poland	& Newton		        & 0.203		& $BVI	$	& 66	& 34 & \ref{TA8},\ref{TA12}	\\
Tenerife-BRT	& Spain		& Schmidt-Cassegrain	& 0.356		& $BVRI_{\rm{C}}$& 23	& 10 & \ref{TA18}	\\
West Challow	& United Kingdom& Schmidt-Cassegrain& 0.356		& $VI_{\rm{C}}$	& 56	& 28 & \ref{TA9}	\\
Wyspa Pucka	    & Poland	& Schmidt-Cassegrain	& 0.203		& $BVR_{\rm{C}}$& 172	& 72 & \ref{TA4}	\\

\hline
IGeoE, Lisboa 	& Portugal	&  Schmidt-Cassegrain &  0.36		& $3840-7360 \AA^{R\sim 483}$ & 4	& 4 &  \ 		\\
Madrid-Ventila	& Spain		&  Ritchey-Chretien		        & 0.2		& $3890-7600 \AA^{R\sim 600}$	& 1	& 1 & \ 	\\
Rozhen		    & Bulgaria	& Ritchey-Chretien*	    & 2.0		& $H_{\alpha}*,H_{\beta}*, Na^{R\sim 16000}$& 10 & 6 & \ 	\\
Rozhen		    & Bulgaria	& Ritchey-Chretien**	& 2.0		& $4000-9000 \AA^{R\sim 30000}$& 2 & 2 & \ 	\\
San Pedro M\'artir & Mexico & Ritchey-Chretien	&	2.1	& $5000-8600 \AA^{R\sim 2000}$ & 3 & 2 & \ \\
West Challow	& United Kingdom& Schmidt-Cassegrain & 0.28		& $3800-7585 \AA^{R\sim  1000}$	& 1	& 1 & \ 	\\
\hline
\end{tabular}
\begin{list}{}{}
\item[{\bf Notes.}] $N_n$ is the number of observed nights. $N_i$ is the number of individual brightness
determinations summed over all the photometric bands. The last column
presents the number of the table with the original data. (*) are spectra made with Coude spectrograph, and (**) are spectra made with ESperRo spectrograph.
\end{list}
\label{telescopes}
\end{table*}

\subsection{Transformation to the standard system}\label{sec_T2SS}

\begin{figure}[h!]
\centering
  \includegraphics[width=7cm]{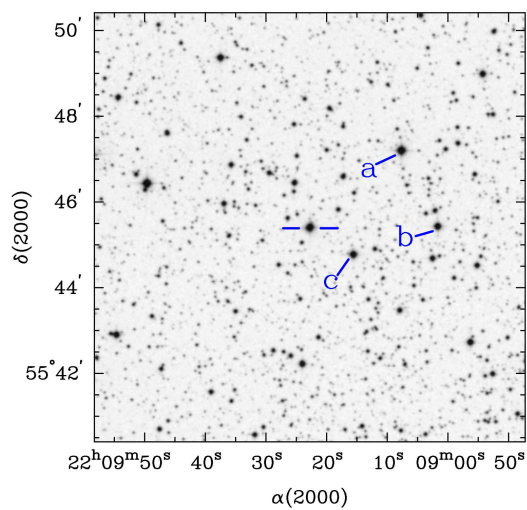}
\caption{$10'\times 10'$ DSS-2-red finding chart for EE Cep.}
  \label{FoV}
\end{figure}

We propose to use the four brightest stars from the Meinuger's sequence
\citep{Mei1975} as comparison and check stars -- all these objects are
close in the sky, within $\sim$3\,\arcmin\, to the position of EE\,Cep
\citep{Gal2014}.  They were marked as "a", "b", "c" and "d" for
BD+55$\degr$\,2690, GSC$-$39732150, BD+55$\degr$\,2691, and GSC$-$39731261,
respectively (Fig.  \ref{FoV}).

Most of the data were expressed by the observers directly as apparent
magnitudes. In these cases we calculated differential magnitude with
respect to the "a" star: $(v - a)$.  In other cases, differential magnitudes
were already given with respect to "a" star.  All data were corrected for
the differences resulting from using different instruments.  We choose two
standard systems.  The photometric data from Suhora Observatory
(Table\,\ref{TA3}) were adopted as a zero-point for $URI$ bands in the
Johnson system.  They covered, with a good density of observational points
in time, the whole time duration of the eclipse as well as a long baseline before and after.
For $BV(RI)_{\rm{C}}$ bands in Johnson-Cousins system, we
chose photometric data from the
University of Athens Observatory (UOAO) (Table\,\ref{TA1}).  The data
consists of 10--50 individual observations every night during the entire
duration of the eclipse (descending and ascending wings, as well as outside
the eclipse).  There are differences between Johnson $RI$ bands and
Johnson-Cousins $(RI)_{\rm{C}}$ bands, so we use different magnitudes of ”a”
star (Table\,\ref{mag}).  For the Johnson system, we used
magnitudes of ”a” star given by \citet{Mik2003}, and for Johnson-Cousins
system we use
magnitudes measured by the University of Athens Observatory (UOAO)
(Table\,\ref{mag}).  We decided to reject differential measurements which
had been done in respect to ”b” and ”c” stars.  According to \citet{Sam2009}
they are suspected to be variable stars.  We mentioned in
\citet{Gal2014} that those stars were monitored in a time period of about 5
years and also during and close to EE Cep eclipse in 2008-2009.  Variability
of ”b” star
wasn't confirmed.  However, we didn't want any unexpected variability from
this star in the course of the eclipse in 2014.

\begin{table}
\caption{Magnitudes of star BD+55$\degr$\,2690 in $UBVRI(RI)_{\rm{C}}$ bands}
\begin{tabular}{ccccccc}
\hline\hline
$U^*$ & $B^{**}$ & $V^{**}$ & $R_{\rm{C}}^{**}$ & $I_{\rm{C}}^{**}$ & $R^{*}$ & $I^{*}$\\
\hline
10.86 & 10.704 & 10.397 & 10.218 & 10.015 & 10.09 & 9.87\\
\hline
\end{tabular}
\label{mag}
\begin{list}{}{}
\item[*] given by \citet{Mik2003}
\item[**] given by the University of Athens Observatory (UOAO)
\end{list}
\end{table}

As it was in the case of previous two campaigns we transform measurements to
one standard system in every band.  The procedure was similar to
\citep{Gal2012} but this time the eclipse was deeper, and we had to take the
effect of the colors into consideration.  The difference between the two systems is not constant as a 
function of depth.  We assumed that the relation between
the shift among standard system and the instrumental one $(v_r - v)$ and
depth of eclipse can be approximated by a linear function.  This relation
takes the form of $f(x) = Ax +
B$, where $x$ is the depth of eclipse measured from the point where $(v - a)
= 0$, and $f = (v_r - v)$ is the shift between standard system $v_r$ and
shifted system $v$.  It was calculated by taking measurements that had been
done at the same time in different systems.  We assumed that observations
which were done at time duration shorter than 0.25 day fulfilled this
condition.  Parameters $A$ and $B$ were estimated by fitting the linear
function to the data.  We obtained instrumental data shifted to zero-point
by 
subtracting calculated shift $v-f(v)$.  We started from the standard system
data set, which had the largest number of averaged measurements.  As a next step, we
added less numerous data sets, ending up to the data sets with two
measurements.  At the end, all photometric data from Tables\,\ref{TA1} to
\ref{TA27} were averaged to one point per night.  Table\,\ref{TA29} contains
this average values with standard deviation and number of measurements used
to average.

\subsection{Spectroscopic data}
\begin{figure*}
\centering
  \includegraphics[width=7cm]{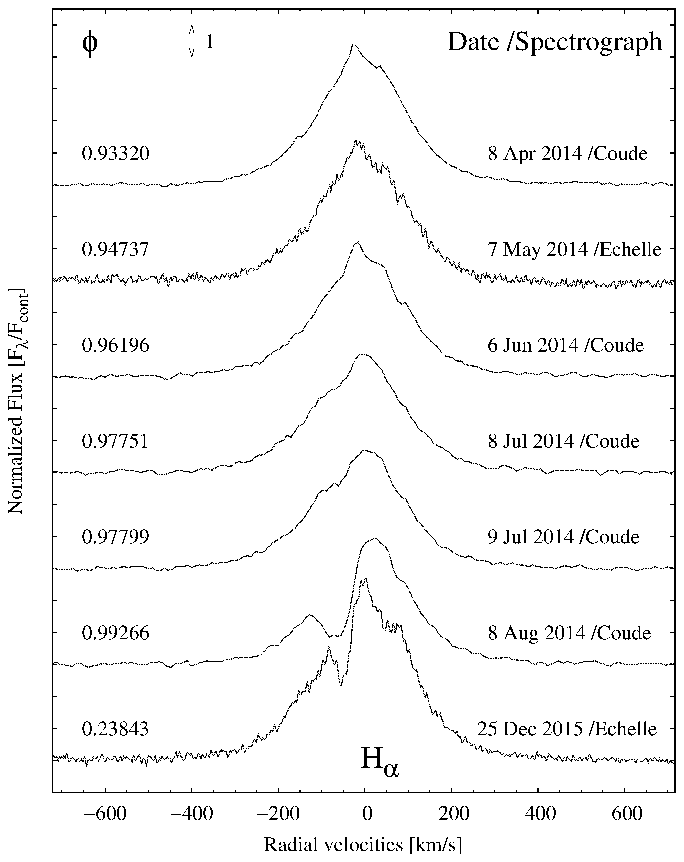}
  \includegraphics[width=7cm]{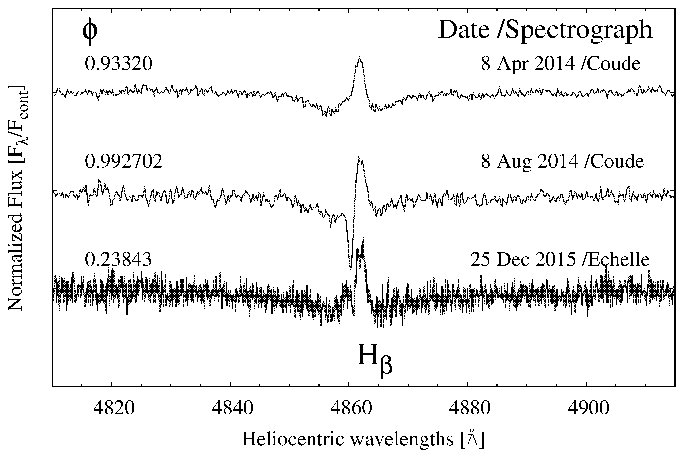}
  \includegraphics[width=7cm]{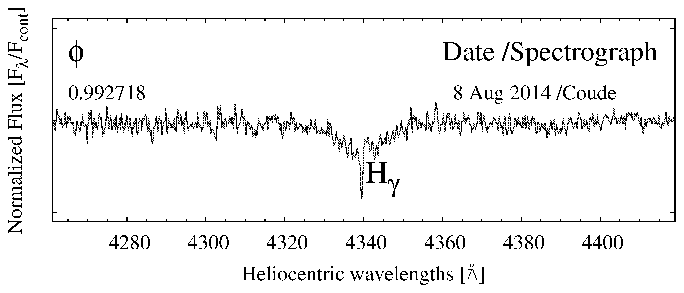}
  \includegraphics[width=7cm]{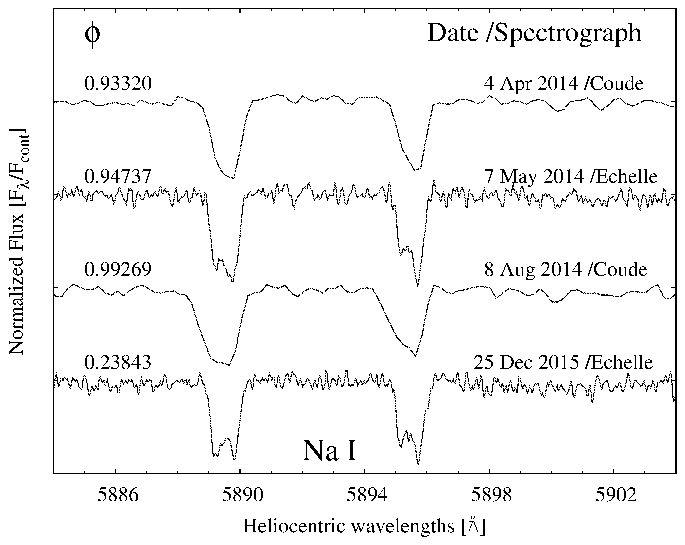}
\caption{Balmer $H_{\alpha}$, $H_{\beta}$ and, $H_{\gamma}$, and
\ion{Na}{I} doublet line profiles in the high-resolution spectra obtained during and
around the eclipse EE\,Cep in 2014 ($E$\,=\,11)}
  \label{spectra-high}
\end{figure*}

High-resolution spectra were collected with two spectrographs (Coud\'e and ESpeRo) operating on the 2-m Ritchey-Chreti\'en telescope at the Rozhen Observatory, Bulgaria. Two spectra ($R \sim 30000$) in range $\sim 4000-9000$ $\AA$ were obtained with ESpeRo echelle spectrograph \citep{Bon2017}. Ten spectra ($R \sim16000$) covering narrow ranges ($\sim 200$\,\AA) were obtained with Coud\'e
spectrograph during 6 nights in the period from April\,4 to August\,6, 2014. Balmer $H_{\alpha}$, $H_{\beta}$ and $H_{\gamma}$ and sodium \ion{Na}{I} doublet line profiles from these spectra are shown in figure \ref{spectra-high}. Nine low-resolution spectra were collected with 4 instruments in range $\sim 4000 - 8600$\,\AA. 
All spectra were heliocentric corrected and normalized to the continuum. The list of spectra and instruments together with some additional information are given in Table\,\ref{telescopes} and all obtained spectra are available as FITS files at the CDS\footnote{Centre de Donn\'ees astronomiques de Strasbourg}.

\section{Results}

\subsection{Photometric Results}

\begin{figure*}
\centering
  \includegraphics[width=16cm]{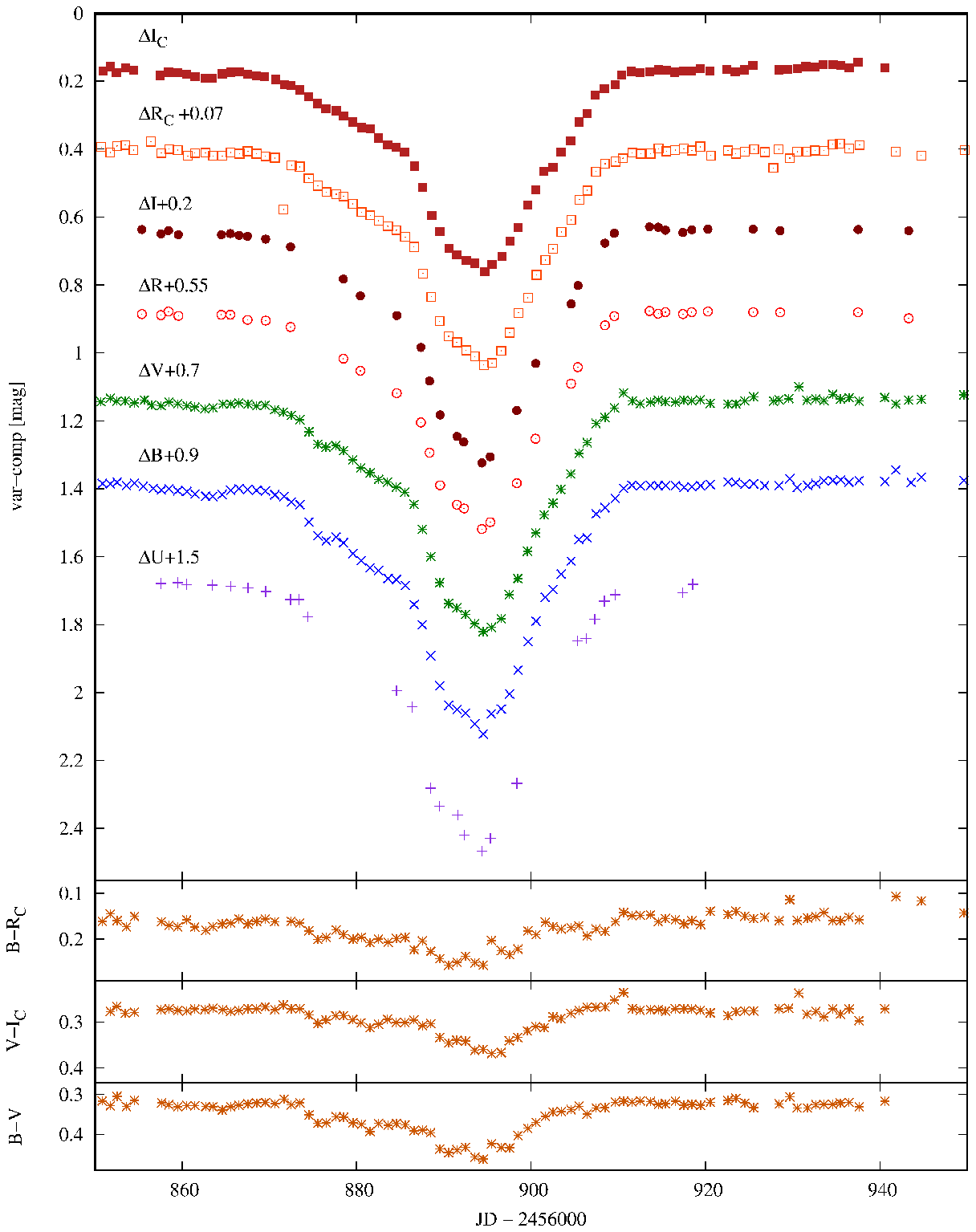}
\caption{Average points from Table\,\ref{TA29} obtained from
photometric measurements of the 2014 eclipse.  $UBVRI(RI)_{\rm{C}}$ light
curves ({\sl Top}) and three colour indices ({\sl Bottom}) are presented.}
  \label{curve}
\end{figure*}

The best coverage with observations was obtained in the $V$ band 
used by every observer.  Coverage in $R_{\rm C}$ and $I_{\rm C}$
is also very dense.  After averaging observations in time
intervals as described in Section\,\ref{sec_T2SS},
 we obtained useful data on 170 individual nights.  The final
$UBVRI(RI)_{\rm{C}}$ light curves are presented in Fig.\ref{curve}.

The minimum of the eclipse took place on $JD = 2456894.0 \pm 0.05$. The timing residual (observations minus calculations -- i.e.  the $(O-C)$ value) with respect to ephemeris from equation \ref{ephemeris} is small $(O-C = 0^d.75)$. The amplitude of the eclipse reached from about $0\fm81$ to $0\fm61$ in $U$ and $I_{\rm{C}}$ bands, respectively. The amplitudes in all bands and the averaged differential magnitudes ($\overline{v - a}$) outside the eclipse (to phase $\sim -0.054$ and from phase $\sim 0.024$) are shown in Table\,\ref{diff}.

\begin{table}
\caption{Average magnitudes of the EE Cep outside the eclipse (to phase
$\sim -0.054$ and from phase $\sim 0.024$), brightnesses in the
minimum of the eclipse and the eclipse depths (differences between
the outside and the minimum values) for seven
particular pass-bands.}
\begin{tabular}{lccccccc}
\hline\hline
Band		& $U$	& $B$	& $V$	& $R_{\rm{C}}$& $I_{\rm{C}}$ & $R$ & $I$ \\
\hline
Outside		& 0.12 	& 0.48	& 0.43	& 0.33 	& 0.15	& 0.31	& 0.41	\\
Minimum		& 0.97	& 1.22	& 1.12	& 0.97	& 0.76	& 0.97	& 1.12	\\
Depth		& 0.83	& 0.75	& 0.71	& 0.65	& 0.62	& 0.66	& 0.71	\\
\hline
\end{tabular}
\label{diff}
\end{table}

\begin{figure*}
\centering
  \includegraphics[width=15cm]{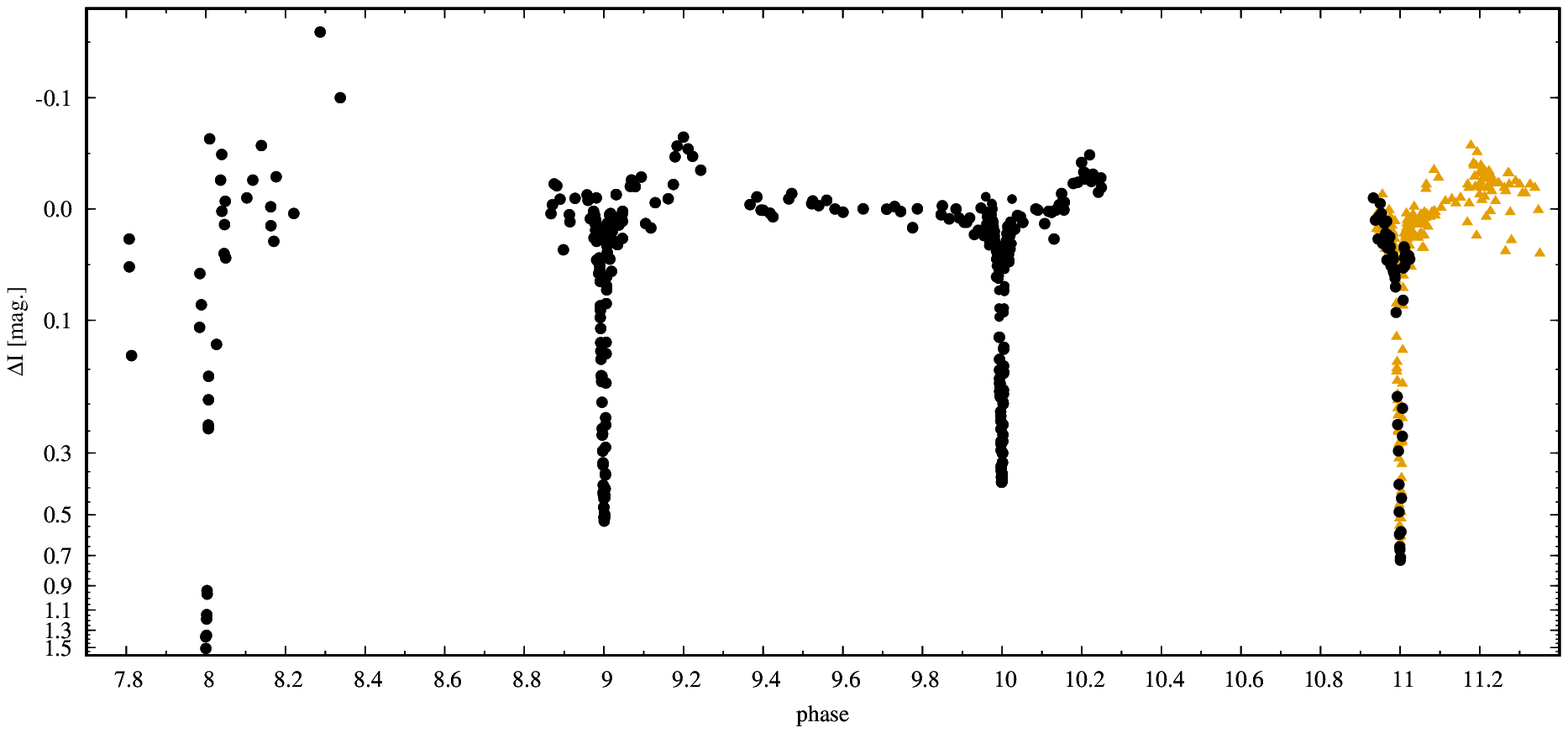}
\caption{Differential $I$ magnitudes of EE Cep during and around of
eclipses at epochs E = 8, 9, 10, and 11. The zero value represents
average brightness outside eclipses (phase 0.4-0.8). Yellow triangles show
the AAVSO $I$-band light curve after correcting for the differences
between photometric systems of particular observers. Magnitudes of epoch E = 11 (black points) are the same as in figure \ref{curve} for $I$-band.}
  \label{Iban}
\end{figure*}

The profiles of the eclipse in all bands are similar to those from 2003 and 2008/09 ($E$\,=\,9 and $E$\,=\,10). In particular, there was a
"bump" at about $\sim9$ days before the photometric minimum during two previous eclipses. During this eclipse event ($E$\,=\,11) it is not so evident in the light curve, like in the all other previous deep eclipses, but it is visible in color indices ($B-I_{\rm{C}}$, $V-I_{\rm{C}}$, and $B-R_{\rm{C}}$) that we calculated (Fig.\ref{curve} -- bottom). Neglecting the second order effects -- short and small amplitude variations -- as observed during both previous campaigns and interpreted as a manifestation of the possible complex, multi-ring structure of the disk \citep{Gal2010}, the observed color indices generally show a two-component profile with a maximum of the "bump" on roughly $\sim$9--10 days before the photometric minimum. The beginning of the fast ingress can be observed in all observed photometric bands. A small ($\sim 0\fm05$) dip is also present, which is visible in $V$ and $B$ light curves at orbital phase $\sim -0.015$ (Figs.\,\ref{spec} and \ref{curve}). It coincides roughly in phase and amplitude (these are subject to change slightly due to changes in the orientation of the disk) with shallow minima, noticed during the eclipses of previous campaigns in $B$ and $V$ light curves, but corresponding events observed previously roughly at a similar time after the mid-eclipse is not clearly detectable. The number of measurements collected in the very near-infrared domain $I$ and $I_{\rm{C}}$ bands) during this campaign turned out unfortunately too small to confirm the repeatable brightening by several hundredths of magnitudes with a maximum at orbital phase $\sim$0.2, which we reported after previous campaigns \citep{Gal2012}. However, the independent campaign by AAVSO collected a relatively large number of measurements in $I$-band. There are significant differences between photometric systems, and the scatter is large, but after correcting for them in the $I$ band light curve for epoch E=11 there also seems to be an apparent maximum at phase $\sim$0.2 (see fig. \ref{Iban}).

Our $JHK_{\rm{S}}$ photometry obtained at phase ($\phi = 0.247$) --
during the time of fall after reaching the maximum brightness --
compared to 2MASS photometry obtained at $\phi = 0.338$ (Table\,\ref{TA28})
shows a slightly increased brightness coherent with the $I$ band, confirming the reality of this phenomenon, and indicating that 
we can expect a somewhat increased amplitude of this effect in
the near infrared region -- with a maximum falling somewhere in the region
of $H$ band.

\subsection{Calculating the new ephemeris}

\begin{figure}
\centering
  \includegraphics[width=9cm]{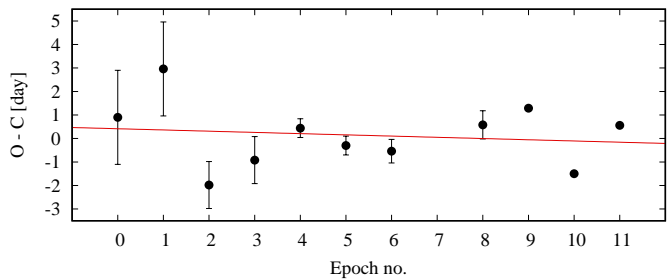}
\caption{ O–C diagram with a linear fit (solid line) to residuals.}
  \label{fig_ephemeris}
\end{figure}

The eclipses of EE Cep are asymmetric, so we couldn't use the standard
methods to calculate the time of
minimum in the light curve. We decided to use the
algorithm described by \citet{Kwe1956}. The same method was used by
\citet{MiGr1999} to calculate the minimum of epoch $E$\,=\,4 eclipse, and
helped to calculate the minima of epochs 2, 5 and 8, and also to calculate the minima of the last three eclipses (epochs 9, 10 and 11). 
Using every known time of minimum for EE\,Cep (Table\,\ref{tab_ephemeris})
we calculated the $O - C$ residuals for the times of minima using the linear
ephemeris $JD(Min)=2434344.1 +2049^d.94 \times E$ by \citet{MiGr1999}.  They
are listed in Table\,\ref{tab_ephemeris} and are marked in
Figure\,\ref{fig_ephemeris}.  The best linear fit to the residuals gives the
new ephemeris:

\begin{equation}
JD(Min)=2434345.16(\pm 2.08)+2049^d.78(\pm 0^d.22) \times E
\label{ephemeris}
\end{equation}

\begin{table}
\caption{Observed moments of minima and the O–C (observations
minus calculations) residuals calculated according to the linear ephemeris
$JD(Min)=2434344.1 +2049^d.94 \times E $ \citep{MiGr1999}}.
\begin{tabular}{lcccl}
\hline
Epoch & Light & JD & O-C & Reference \\
\hline

 0 & mpg & $2434345.0  \pm 2.0$  & 0.90   & [1]  \\ 
 1 & mpg & $2436397.0  \pm 2.0$  & 2.96   & [1]  \\ 
 2 & mpg & $2438442.0  \pm 1.0$  & -1.98  & [2]  \\ 
 3 & mpg & $2440493.0  \pm 1.0$  & -0.92  & [3]  \\ 
 4 & V   & $2442544.3  \pm 0.4$  & 0.44   & [2]  \\ 
 5 & V   & $2444593.5  \pm 0.4$  & -0.30  & [2]  \\ 
 6 & V   & $2446643.2  \pm 0.5$  & -0.54  & [4]  \\ 
 7 & V   & $2448691.87 \pm 0.89$ & -1.78  & [5]  \\ 
 8 & V   & $2450744.2  \pm 0.6$  & 0.58   & [2]  \\ 
 9 & V   & $2452794.85 \pm 0.14$ & 1.29   & this work \\ 
 10& V   & $2454842.0  \pm 0.1$ & -1.50  & this work \\ 
 11& V   & $2456894.0  \pm 0.2$ & 0.56   & this work \\ 
\hline
\end{tabular}
\begin{list}{}{}
\item[References:] $^{[1]}$\,\citet{Mei1973}; $^{[2]}$\,\citet{MiGr1999}; $^{[3]}$\,\citet{Bal1975}; $^{[4]}$\,\citet{DiL1988}; $^{[5]}$\,\citet{Hal1992}
\end{list}
\label{tab_ephemeris}
\end{table}

\subsection{Spectroscopic Results}

\begin{figure*}{}
\centering
  \includegraphics[width=12cm]{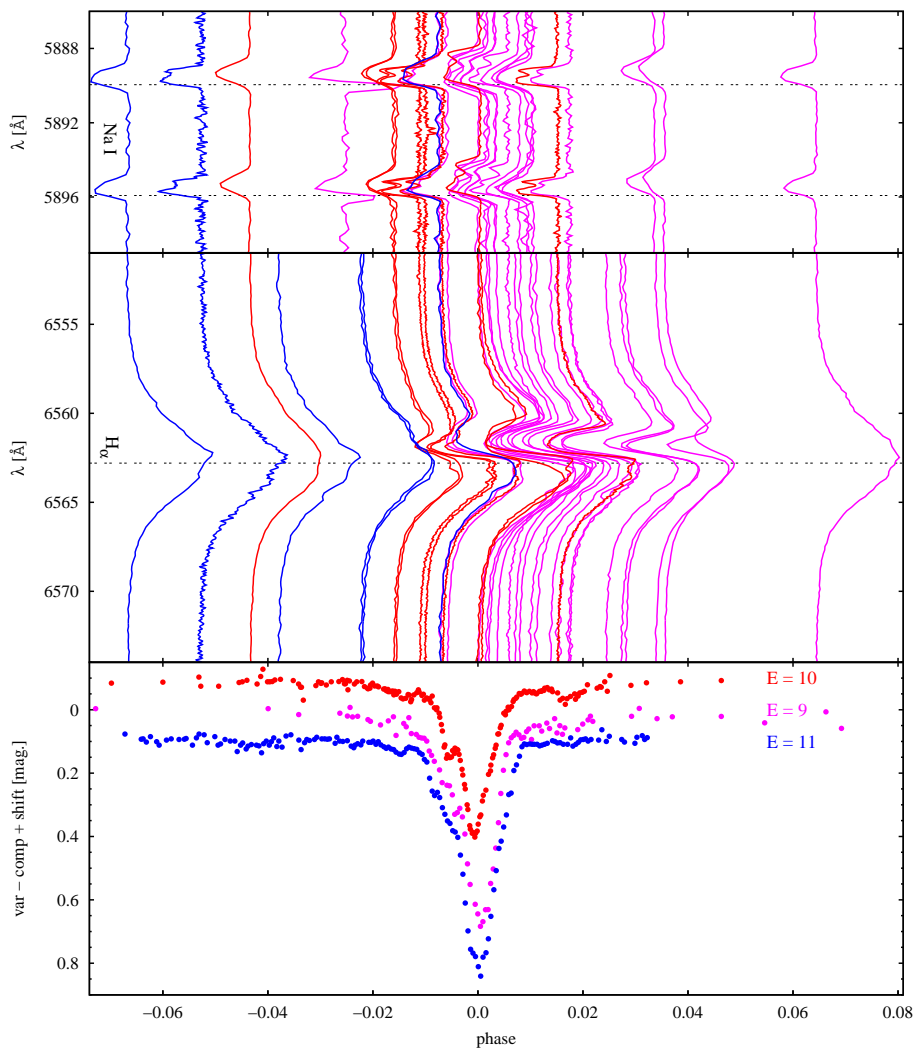}
\caption{Evolution of the \ion{Na}{I} line profiles $(Top)$,
$H_{\alpha}$ line profiles $(Middle)$  and $B$ light curves $(Bottom)$
during and around of eclipses at epochs $E$ = 9, 10 and 11.  Data are folded
according to ephemeris by equation \ref{ephemeris}.}

  \label{spec}
\end{figure*}

We have not observed any significant changes in the $H_{\alpha}$ line up to
July 8, and 9, 2014, when signs of absorption components began to appear
on the blue wings of the emission component (Figs \ref{spectra-high} and \ref{spec}). 
It coincides with the time when the small dip starts to develop in the light
curves, which signifies the beginning of the photometric eclipse.  The new
precise photometric data with dense time coverage reveals more
subtle changes and more accurate determination of the eclipse times and
their total duration.  The shallow eclipses last up to 2.5 – 3.5 months and
there is now less difference with results obtained from spectroscopic
observations.  The total duration times estimated from the old photometric
data \citep{Gra2003} seem to be underestimated.  The Balmer $H_{\alpha}$,
$H_{\beta}$ and $H_{\gamma}$ profiles in the spectra obtained deep in the
eclipse (Aug\,8,\,2014) show deep absorption component on their blue wings. 
Interestingly, $H_{\alpha}$ and $H_{\beta}$ lines in the spectra obtained on
Dec\,25, 2015, also show absorption component -- this is the orbital phase
$\sim$0.24 during the brightening event (soon after its maximum) in the
$I$-band.  The blue-shifted absorption components are visible also in
\ion{Na}{I} doublet observed during this phase. Sodium
develops absorption before and after the external photometric
contacts of the eclipses (eg.  spectrum on May\,7, 2014). Similar behaviour
of the sodium and potassium lines was observed in $\varepsilon$\,Aur system
(see eg.  \citet{Lea2012}; \citet{Tom2012}). Several low-resolution
(R\,$\sim$\,500\,--\,1000)  spectra listed in
Table\,\ref{telescopes} generally confirm the evolution of the
EE Cep spectrum shown by \citet{Boy2014}.

\section{Precessing Be star model}

\begin{figure}{}
\centering
  \includegraphics[width=8.5cm]{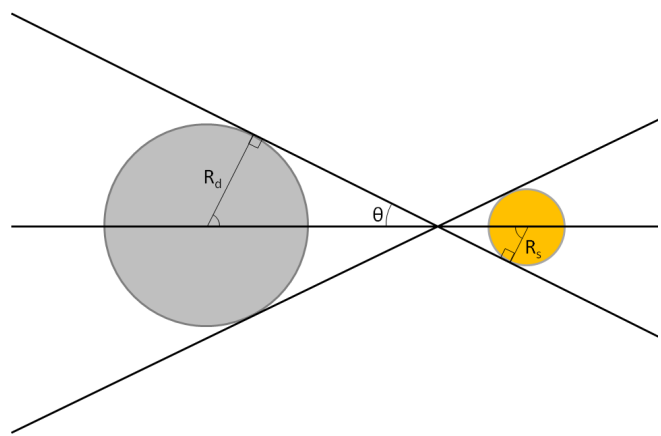}
\caption{Simple assumptions about geometry of the EE Cep system.}
  \label{geometry}
\end{figure}

Analysis of the  photometric data of the last
campaigns together with the changes in the spectral line profiles proves
that the eclipses have longer total duration, even up to $\sim$3.5 months,
than postulated on the basis of observations of previous eclipses. The last
eclipse reached a depth of only $\sim$0$\fm$75 mag in $B$, much shallower than expected $\sim2\fm0$ according to \citet{Gal2012} model. One possible reason can be that the assumptions made to the disk precession model were too simple.
In particular: ({\sl i}) circularity of the orbit, and ({\sl ii}) very
simplified, disk density profile ($\sim r^{-2}$).  Moreover,
({\sl iii}) the disk diameter was adopted quite arbitrarily as not larger than $D\sim$150\,R\sun.
This could be justified by making simple assumptions about the system geometry shown on figure \ref{geometry}. The disk size can be calculated as a function of component separation $a$, companion radius $R_S$, and viewing angle 
$\theta$. The viewing angle corresponds to half the duration of the eclipse which is about 3 months, thus $\theta \simeq 0.056$. We assumed that disk mass is much lower than the Be star mass (which is $\sim 8 M_{\odot}$ according to \citet{Gal2012} as well as $R_S \simeq 9R_{\odot}$). Taking this into consideration the component separation determined by Keplerian law is $a = 1360 R_{\odot}$, which gives the disk diamater about $370 R_{\odot}$. However this estimation doesn't take into consideration impact parameter $D$, inclination and eccentricity of the system.   

In future work, we are going to take into consideration another
model that could explain changes in eclipse depths. At the moment the
hypothesis of eclipses caused by a disk still offers the best way to explain
the grey character of all eclipses and that the secondary component is still
 eluding detection. It is also a good explanation for small changes in
the light curve in the early stage of the eclipses.

\begin{figure*}{}
\centering
  \includegraphics[width=15cm]{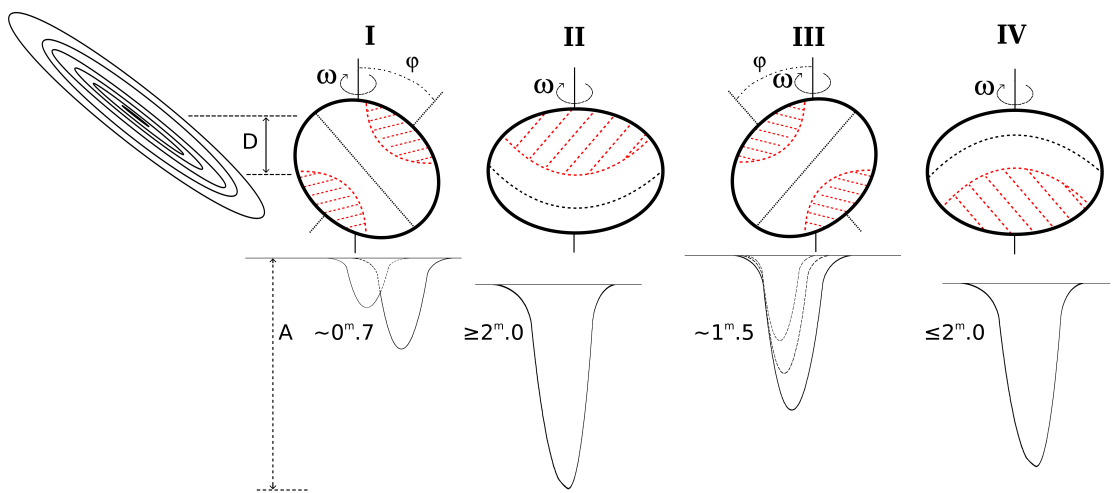}
\caption{Precessing Be star model explaining changes in the depths of the
eclipses}

  \label{new_model}
\end{figure*}

We want to propose a model in which the Be star is precessing instead of the
disk. In this case, the changes in the eclipse depth would be
caused by eclipsing the hot spots on the poles of the star as shown in the scheme on figure \ref{new_model}.
We can distinguish four special orientations in the geometry of the
system, where the angle of the disk and
impact parameter $D$ is the same.  Only the inclination of the star
$\varphi$ is changing due to precession which would cause different amplitude $A$ of the eclipses.

\begin{list}{}{}
\item I. Two clearly separate, relatively shallow minima with depth $\sim 0^m.7$ (epochs 9, 10, 11). First is the occultation of the bottom pole, then top one.
\item II. Only one, the visible hot pole is being highly eclipsed  (epochs 0, 1, 2). The hot spot is close to the disk center. Deep ($A \gtrsim 2^m.0$) eclipses are observed.
\item III. Overlapping minima from two poles that are being eclipsed almost simultaneously (epochs 4, 5, 6). The hot spot caused by the bottom pole experiences a weaker eclipse ($A \sim 1^m.5$) through the outer, more transparent parts of the disk. Intermediate depth eclipses are observed.
\item IV. The hot spot, further from the disk center is eclipsed (epoch 8). Similar to case II but the eclipse is less deep $A \lesssim 2^m.0$.
\end{list}

Changes in the inclination of Be star (precession) may cause small changes in its
brightness and colors due to the visibility of hot polar regions. 
Line broadening may also change during the precession cycle.  Our future
studies will concentrate on veryfing this model.

\section{Concluding remarks}

In this work, we present the processed photometric and spectroscopic data
that have been obtained during the recent eclipse season
of EE~Cep in 2014 ($E$\,=\,11). We can formulate several conclusions as to possibly the most important observational features of the eclipses which should be taken into account:

\begin{list}{}{}
\item There are small effects detectable in the light and color curves which
could be related to a possible complex multi-ring structure of the disk.
\citet{Gal2010} speculated that possible planets could be responsible for
the formation of the gaps in the disk. However, this adds an extra complication
to the model and such small second-order effects should be neglected
in the modeling of the precession.\\

\item There are indications that the orbit is significantly eccentric.
Variations out of the eclipses noted in $I$-band light curves at orbital
phase $\sim0.2$ \citep[][and this paper]{Gal2012} we believe could be
related to proximity effects, when components are approaching each other
close to periastron passage. In that case, the acceleration in the orbital
motion may be an additional reason for the observed eclipse asymmetry and
this complication may have to be taken into account in the
model.\\

\item The most serious problem for developing the unified model of all
eclipses are two events at epochs $E$\,=\,3 and $E$\,=\,8 which are characterised with extremely different depths from neighbouring
eclipses. In the case of the disk precession model this effect
could be reproduced by realization of the scenario in which two effects compete
when the disk is seen nearly edge-on: ({\sl i}) the rapid change in the
optical depth of the disk, and ({\sl ii}) the
change in size of the disk projection which makes changes in the extent of
obscuration \citep[see][section 4.4]{Gal2012}.  The larger disk diameter
should enhance  the efficiency of
this mechanism because it will imply a smaller inclination of the precession axis.
In the alternative option of the star precession some other mechanism explaining such quick changes will have to be proposed -- eg.  it can be considered to what extent the
slope of the star's precession axis could be responsible for this.

\end{list}

Applying new modelling techniques might bring a breakthrough.
For example, any possibility to know parameters of the orbit from measurements of
the radial velocity variations (very difficult for such a rapidly rotating,
hot star) and/or if possible the movement of the cold component from
infrared interferometry, may be crucial for constraining the parameters and
understanding of this system. Our still scant infrared data ($IJHK$--bands)
seem to indicate the possibility to detect the cold component through systematic
infrared monitoring during the whole orbital cycle. 

\section{Call for observations}

We announce another observation campaign of EE Cep which will be held in
2020.  The ingress will start at about
7th March (JD 2458916.4), whereas the end of egress will be observed around 21st April (JD
2458960.5) with mid-eclipse around 3rd April (JD 2458943.19).

Due to the duration of the previous eclipses, it is recommended to start the
observations at least two months before the mid-eclipse and
continue for four months.

Attention should be paid around 20th May 2021 (JD 2459354.7) due to the fact
that in previous epochs flux increase in $I$ band was observed around these
days.  This phenomenon will start around 25th October, 2020 (JD 2459148.1)
and end around 9 December, 2021 (JD 2459558.2).

It is recommended to observe in the Johnson-Cousin system with an
accuracy of 0$\fm$01 or higher.  In case of $J$,$H$,$K$ infrared bands, all
possible observations will be useful.

As a comparison star, we suggest star ``$a$", BD+55$^{\circ}$2690, but we
also encourage to observe stars ``$b$", ``$c$", and ``$d$" for
GSC-3973\,2150,
BD+55$^{\circ}$2691, and GSC-3973\,1261, respectively.  

Stars ``$b$" and ``$c$" are designated as New Suspected Variables in the General Catalog of
Variable Stars (Samus et al.\,2009).  Observations should verify this possibility.

We are interested in spectroscopic observations with a resolution of R
$\sim 10$\,$000$ or higher.  Observing lines H$_{\alpha}$, H$_{\beta}$ and H$_{\gamma}$, and
\ion{Na}{I} doublet for the analysis of spectral profiles is encouraged. 
In the case of spectrographs
with low resolutions only observations calibrated in flux will become useful.

Support necessary for maintaining the observations may be found at:
http://sites.google.com/site/eecep2020campaign/.
Observers interested in taking part in the campaign are requested
to contact Dariusz Kubicki at kubickid@gmail.com.

\begin{acknowledgements}
This study has been partly supported by the Polish National Science
Centre grant No DEC-2015/19/D/ST9/02974.  This paper is partly a result
of the exchange and joint research project Spectral and photometric studies
of variable stars between Polish and Bulgarian Academies of Sciences. A large 
number of photometric data on EE~Cep were collected during the photometric 
monitoring observations with the robotic and remotely controlled observatory 
at the University of Athens Observatory - UOAO \citep{Gaz2016}.
\end{acknowledgements}

%
%

\bibliographystyle{aa} 

%

%

\begin{appendix} 

\section{Online photometric and spectroscopic data}\label{AppendixT}

\addtocounter{table}{1}
\longtab[1]{
\begin{landscape}
\begin{longtable}{|lrrrrrrrrrrrrrrrr|}
   \caption{\label{TA1} Photometry obtained at Athens Observatory (Greece) with standard
$BV(RI)_{\rm{C}}$ (Bessell) filters during and near the 2014 eclipse (E = 11).  The
0.4 m reflector telescope with an SBIG ST10 XME CCD camera was used.
Differential magnitudes are given with respect to $BD+55$\,2690.  Each point
is the mean value obtained from 3 h frames, together with the corresponding
standard deviations and number of points that were used to average. The columns
labeled 'HJD+' denote the fraction of the
day.}\\
\hline\hline
$HJD$ & $HJD+$ & $\Delta B$ & $\sigma_{B}$ & $N_B$ & $HJD+$ & $\Delta V$ & $\sigma_{V}$ & $N_V$ & $HJD+$ & $\Delta R_C$ & $\sigma_{R_C}$ & $N_{R_C}$ & $HJD+$ & $\Delta I_C$ & $\sigma{I_C}$ & $N_{I_C}$\\
\hline
\endfirsthead
\caption{continued.}\\
\hline\hline
$HJD$ & $HJD+$ & $\Delta B$ & $\sigma_{B}$ & $N_B$ & $HJD+$ & $\Delta V$ & $\sigma_{V}$ & $N_V$ & $HJD+$ & $\Delta R_C$ & $\sigma_{R_C}$ & $N_{R_C}$ & $HJD+$ & $\Delta I_C$ & $\sigma_{I_C}$ & $N_{I_C}$\\
\hline
\endhead
\hline
\endfoot
2456839 & .37866 & 0.485 & 0.006 & 10 & .37978 & 0.432 & 0.008 & 10 & .38043 & 0.323 & 0.007 & 10 & .38109 & 0.161 & 0.008 & 10\\
2456859 & .59785 & 0.502 & 0.007 & 13 & .59836 & 0.448 & 0.008 & 13 & .59876 & 0.344 & 0.006 & 13 & .59723 & 0.176 & 0.009 & 15\\
2456860 & .59779 & 0.502 & 0.004 & 8 & .59831 & 0.456 & 0.007 & 8 & .59870 & 0.339 & 0.006 & 8 & .59912 & 0.169 & 0.008 & 8\\
2456864 & .57769 & 0.503 & 0.006 & 20 & .57878 & 0.451 & 0.008 & 20 & .57942 & 0.342 & 0.007 & 20 & .58007 & 0.177 & 0.011 & 20\\
2456865 & .57204 & 0.496 & 0.006 & 21 & .57477 & 0.446 & 0.006 & 22 & .57542 & 0.342 & 0.004 & 22 & .57608 & 0.173 & 0.005 & 22\\
2456866 & .56753 & 0.500 & 0.008 & 25 & .56865 & 0.449 & 0.006 & 25 & .56930 & 0.345 & 0.008 & 25 & .56995 & 0.175 & 0.005 & 25\\
2456867 & .56976 & 0.496 & 0.005 & 25 & .57088 & 0.446 & 0.006 & 25 & .57153 & 0.344 & 0.006 & 25 & .57218 & 0.175 & 0.005 & 25\\
2456868 & .56914 & 0.502 & 0.003 & 26 & .57025 & 0.453 & 0.004 & 26 & .57091 & 0.350 & 0.004 & 26 & .57156 & 0.182 & 0.004 & 26\\
\end{longtable}
\begin{list}{}{}
\item[Observers:] K. Gazeas.
\end{list}
\end{landscape}
}

\addtocounter{table}{2}
\longtab[2]{
\begin{landscape}
\begin{longtable}{|l@{\hskip 3mm}r@{\hskip 3mm}r@{\hskip 3mm}r@{\hskip 3mm}r@{\hskip 3mm}r@{\hskip 3mm}r@{\hskip 3mm}r@{\hskip 3mm}r@{\hskip 3mm}r@{\hskip 3mm}r@{\hskip 3mm}r@{\hskip 3mm}r@{\hskip 3mm}r@{\hskip 3mm}r@{\hskip 3mm}r@{\hskip 3mm}r@{\hskip 3mm}r@{\hskip 3mm}r|}
   \caption{\label{TA2} $BV(RI)_{\rm{C}}$ photometry obtained at Sonoita
(USA) Observatory during and near the 2014 eclipse (E = 11).  The  0.5
Folded Newtonian telescope with an SBIG STL-1001E CCD camera was used.
Comparisons were made against three stars: $BD+55$\,2690, $GSC-3973 2150$,
$BD+55$\,2691.  The apparent magnitudes are also given with respect to
$BD+55$\,2690 (calculated with the adoption of $BV(RI)_{\rm{C}}$ magnitudes
of $BD+55$\,2690 given by Athens Observatory).  The columns labeled 'JD+'
denote the fraction of the day.}\\

\hline\hline
JD & $JD+$ &$B$&$\Delta B$& $JD+$ &$V$&$\Delta V$& $JD+$ &$R_C$&$\Delta R_C$& $JD+$ &$I_C$&$\Delta I_C$& $JD+$ & $B-Rc$ & $JD+$ & $B-V$ & $JD+$ & $V-Rc$\\
\hline
\endfirsthead
\caption{continued.}\\
\hline\hline
JD &$ JD+ $&$B$&$\Delta B$&$ JD+ $&$V$&$\Delta V$&$ JD+ $&$R_C$&$\Delta R_C$&$ JD+ $&$I_C$&$\Delta I_C$&$ JD+ $& B-Rc &$ JD+ $& B-V &$ JD+ $& V-Rc\\
\hline
\endhead
\hline
\endfoot

2456761 & .98694 & 11.162 & 0.458 & .98750 & 10.800 & 0.403 & ---   & ---    & ---    & ---    & ---    & --- & --- & --- & --- & ---& --- & ---\\
2456763 & .98641 & 11.171 & 0.467 & .98699 & 10.803 & 0.406 & ---   & ---    & ---    & ---    & ---    & --- & --- & --- & --- & ---& --- & ---\\
2456767 & .98415 & 11.170 & 0.466 & .98472 & 10.806 & 0.409 & ---   & ---    & ---    & ---    & ---    & ---   & ---    & ---   & ---    & ---  & ---    & ---\\
2456768 & .97720 & 11.181 & 0.477 & .97778 & 10.799 & 0.402 & .97926 & 10.529 & 0.311 & ---    & ---    & ---   & .97823 & 0.652 & .98722 & 0.362& .97852 & 0.270\\
2456771 & .98258 & 11.169 & 0.465 & .98315 & 10.808 & 0.411 & .98356 & 10.542 & 0.324 & .98399 & 10.196 & 0.181 & .98307 & 0.627 & .98670 & 0.368& .98336 & 0.266\\
2456776 & .99179 & 11.184 & 0.480 & .99235 & 10.791 & 0.394 & .99277 & 10.526 & 0.308 & .99318 & 10.182 & 0.167 & .99228 & 0.658 & .98443 & 0.364& .99256 & 0.265\\
2456777 & .96637 & 11.192 & 0.488 & .96589 & 10.804 & 0.407 & .96631 & 10.531 & 0.313 & .96673 & 10.189 & 0.174 & .96634 & 0.661 & .97749 & 0.382& .96610 & 0.273\\
2456778 & .96483 & 11.163 & 0.459 & .96540 & 10.810 & 0.413 & .96580 & 10.540 & 0.322 & .96623 & 10.186 & 0.171 & .96532 & 0.623 & .98287 & 0.361& .96560 & 0.270\\
\end{longtable}

\begin{list}{}{}
\item[Observers:] B. Steals.
\end{list}
\end{landscape}

}


\addtocounter{table}{3}
\longtab[3]{
\tiny{
\begin{landscape}
\begin{longtable}{|lrrrrrrrrrrrrrrrr|}
   \caption{\label{TA3} Photometry in standard $UBVRI$ filters obtained at
Suchora Observatory (Poland) during and near the 2014 eclipse (E = 11).  The
0.6 m Cassegrain telescope with a CCD camera was used.  Differential
magnitudes are given with respect to $BD+55$\,2690.  The columns labeled
'HJD+' denote the fraction part of the day.}\\

\hline\hline
$HJD $&$HJD+$&$\Delta U$&$HJD+$&$\Delta B$&$HJD+$&$\Delta V$&$HJD+$&$\Delta R$&$HJD+$&$\Delta I$&$ JD+ $&$B-R_C$ &$ JD+ $&$B-V$ &$ JD+ $&$V-R_C$\\
\hline
\endfirsthead
\caption{continued.}\\
\hline\hline
$HJD $&$HJD+$&$\Delta U$&$HJD+$&$\Delta B$&$HJD+$&$\Delta V$&$HJD+$&$\Delta R$&$HJD+$&$\Delta I$&$ JD+ $&$B-R_C$ &$ JD+ $&$B-V$ &$ JD+ $&$V-R_C$\\
\hline
\endhead
\hline
\endfoot

2456755 & --- & --- & .59526 & 0.437 & .59546 & 0.385 & .59023 & 0.290 & .59054 & 0.118 & .59275 & 0.147 & .59536 & 0.052 & .59285 & 0.095\\
2456765 & --- & --- & .52500 & 0.459 & .52521 & 0.403 & .52535 & 0.288 & .52547 & 0.131 & .52517 & 0.171 & .52511 & 0.056 & .52528 & 0.115\\
2456765 & --- & --- & --- & --- & --- & --- & --- & --- & --- & --- & .54358 & 0.159 & .54115 & 0.053 & .54369 & 0.103\\
2456765 & --- & --- & --- & --- & --- & --- & --- & --- & --- & --- & .56264 & 0.173 & .56258 & 0.058 & .54133 & 0.118\\
2456765 & .59236 & 0.124 & .59994 & 0.461 & .55730 & 0.406 & .56217 & 0.300 & .56483 & 0.131 & .58105 & 0.161 & .57862 & 0.055 & .55974 & 0.106\\
2456779 & --- & --- & .48204 & 0.466 & .48454 & 0.422 & .48854 & 0.316 & .48705 & 0.138 & .48529 & 0.150 & .48329 & 0.044 & .48654 & 0.106\\
2456783 & --- & --- & .51704 & 0.455 & .52378 & 0.402 & .52390 & 0.305 & .52399 & 0.129 & .52047 & 0.150 & .52041 & 0.053 & .52384 & 0.097\\
2456784 & --- & --- & .55680 & 0.452 & .55566 & 0.402 & .55288 & 0.301 & .55302 & 0.126 & .55484 & 0.151 & .55623 & 0.050 & .55427 & 0.101\\
\end{longtable}

\begin{list}{}{}
\item[Observers:] S. Zola.
\end{list}
\end{landscape}
}
}

\addtocounter{table}{4}
\longtab[4]{
\begin{landscape}
\begin{longtable}{|lrrrrrrrrrrrrrrr|}
   \caption{\label{TA4} $UBV(RI)_{\rm{C}}$ photometry obtained at Wyspa
Pucka Observatory (Poland) during and near the 2014 eclipse (E = 11).  The
0.203 Schmidt-Cassegrain telescope with an Audine CCD camera was used.  The
apparent magnitudes are also given with respect to $BD+55$\,2690 (calculated
with the adoption of $BV(RI)_{\rm{C}}$ magnitudes of $BD+55$\,2690 given by
Athens Observatory.  Comparison was made against star $BD+55$\,2690.  The
columns labeled 'JD+' denote the fraction of the day.}\\

\hline\hline
JD &$ JD+ $&$B$&$\Delta B$&$ JD+ $&$V$&$\Delta V$&$ JD+ $&$R_C$&$\Delta R_C$&$ JD+ $&$B-R_C$ &$ JD+ $&$B-V$ &$ JD+ $&$V-R_C$\\
\hline
\endfirsthead
\caption{continued.}\\
\hline\hline
JD &$ JD+ $&$B$&$\Delta B$&$ JD+ $&$V$&$\Delta V$&$ JD+ $&$R_C$&$\Delta R_C$&$ JD+ $&$B-R_C$ &$ JD+ $&$B-V$ &$ JD+ $&$V-R_C$\\
\hline
\endhead
\hline
\endfoot

2456840 & --- & --- & --- & --- & --- & --- & .45061 & 10.532 & 0.314 & --- & --- & --- & --- & --- & ---\\
2456841 & --- & --- & --- & --- & --- & --- & .45282 & 10.529 & 0.311 & --- & --- & --- & --- & --- & ---\\
2456842 & --- & --- & --- & --- & --- & --- & .46125 & 10.529 & 0.311 & --- & --- & --- & --- & --- & ---\\
2456843 & --- & --- & --- & --- & --- & --- & .47591 & 10.517 & 0.299 & --- & --- & --- & --- & --- & ---\\
2456844 & --- & --- & --- & --- & --- & --- & .37384 & 10.522 & 0.304 & --- & --- & --- & --- & --- & ---\\
2456845 & --- & --- & --- & --- & --- & --- & .45994 & 10.519 & 0.301 & --- & --- & --- & --- & --- & ---\\
2456846 & --- & --- & --- & --- & --- & --- & .40497 & 10.514 & 0.296 & --- & --- & --- & --- & --- & ---\\
2456848 & --- & --- & --- & --- & --- & --- & .47192 & 10.514 & 0.296 & --- & --- & --- & --- & --- & ---\\
\end{longtable}

\begin{list}{}{}
\item[Observers:] A. Arminski.
\end{list}
\end{landscape}
}
\addtocounter{table}{5}
\longtab[5]{
\begin{landscape}
\begin{longtable}{|lrrrrrrrrrrrrrrrrrr|}
   \caption{\label{TA5} $BV(RI)_{\rm{C}}$ photometry obtained at
Bia{\l}k\'ow (Poland) Observatory during and near
the 2014 eclipse (E = 11). The 0.6 Cassegrain telescope with aANDOR DW-432 BV
CCD camera was used. Comparisons were made against star: $BD+55$\,2690. The
apparent magnitudes are also given with respect to $BD+55$\,2690 (calculated with
the adoption of $BV(RI)_{\rm{C}}$ magnitudes of $BD+55$\,2690 given by Athens
Observatory). Comparison was made against star $BD+55$\,2690. The columns
labeled 'JD+' denote the fraction of the day.}\\
\hline\hline
JD &$ JD+ $&$B$&$\Delta B$&$ JD+ $&$V$&$\Delta V$&$ JD+ $&$R_C$&$\Delta R_C$&$ JD+ $&$I_C$&$\Delta I_C$&$ JD+ $&$B-R_C$ &$ JD+ $&$B-V$ &$ JD+ $&$V-R_C$\\
\hline
\endfirsthead
\caption{continued.}\\
\hline\hline
JD &$ JD+ $&$B$&$\Delta B$&$ JD+ $&$V$&$\Delta V$&$ JD+ $&$R_C$&$\Delta R_C$&$ JD+ $&$I_C$&$\Delta I_C$&$ JD+ $&$B-R_C$ &$ JD+ $&$B-V$ &$ JD+ $&$V-R_C$\\
\hline
\endhead
\hline
\endfoot
2456798 & .53712 & 11.184 & 0.480 & .53880 & 10.828 & 0.431 & .53867 & 10.534 & 0.316 & .54226 & 10.164 & 0.149 & .53790 & 0.650 & .53796 & 0.356 & .53874 & 0.294\\
2456799 & .54637 & 11.188 & 0.484 & .54523 & 10.831 & 0.434 & .54727 & 10.538 & 0.320 & .54862 & 10.179 & 0.164 & .54682 & 0.650 & .54580 & 0.357 & .54625 & 0.293\\
2456800 & .54626 & 11.182 & 0.478 & .54885 & 10.825 & 0.428 & .54997 & 10.530 & 0.312 & .55065 & 10.167 & 0.152 & .54812 & 0.652 & .54756 & 0.357 & .54941 & 0.295\\
2456815 & .52560 & 11.181 & 0.477 & .52870 & 10.826 & 0.429 & .53052 & 10.532 & 0.314 & .53198 & 10.168 & 0.153 & .52806 & 0.649 & .52715 & 0.355 & .52961 & 0.294\\
2456827 & .52511 & 11.181 & 0.477 & .52822 & 10.828 & 0.431 & .53112 & 10.532 & 0.314 & .53136 & 10.167 & 0.152 & .52811 & 0.649 & .52667 & 0.353 & .52967 & 0.296\\
2456845 & .51170 & 11.194 & 0.490 & .51466 & 10.839 & 0.442 & .51638 & 10.540 & 0.322 & .51758 & 10.183 & 0.168 & .51404 & 0.654 & .51318 & 0.355 & .51552 & 0.299\\
2456857 & .44986 & 11.209 & 0.505 & .45340 & 10.850 & 0.453 & .45509 & 10.552 & 0.334 & .45627 & 10.190 & 0.175 & .45248 & 0.657 & .45163 & 0.359 & .45425 & 0.298\\
2456858 & .42078 & 11.206 & 0.502 & .42257 & 10.846 & 0.449 & .42496 & 10.550 & 0.332 & .42782 & 10.183 & 0.168 & .42287 & 0.656 & .42167 & 0.360 & .42376 & 0.296\\
\end{longtable}

\begin{list}{}{}
\item[Observers:] Z. Kolaczkowski, D. Mo"zdzierski, E. Zahajkiewicz, P. Bru"s, A. Pigulski
\end{list}
\end{landscape}
}

\addtocounter{table}{6}
\longtab[6]{
\begin{landscape}
\begin{longtable}{|l@{\hskip 3mm}r@{\hskip 3mm}r@{\hskip 3mm}r@{\hskip 3mm}r@{\hskip 3mm}r@{\hskip 3mm}r@{\hskip 3mm}r@{\hskip 3mm}r@{\hskip 3mm}r@{\hskip 3mm}r@{\hskip 3mm}r@{\hskip 3mm}r@{\hskip 3mm}r@{\hskip 3mm}r@{\hskip 3mm}r@{\hskip 3mm}r@{\hskip 3mm}r@{\hskip 3mm}r@{\hskip 3mm}r@{\hskip 3mm}r@{\hskip 3mm}r|}
   \caption{\label{TA6} $UBV(RI)_{\rm{C}}$ photometry obtained at Slovak Academy (Slovakia) Observatory during
and near the 2014 eclipse (E = 11). The 0.6 Cassegrain telescope with a
Moravian Instruments G4-9000 CCD camera was used. Comparisons were made against
three stars: $BD+55$\,2690, $GSC-3973$\,2150, $BD+55$\,2691. The apparent magnitudes are
also given with respect to $BD+55$\,2690 (calculated with the adoption of $UBV(RI)_{\rm{C}}$
magnitudes of $BD+55$\,2690 given by Athens Observatory). Comparison was made
against star $BD+55$,2690. The columns labeled 'JD+' denote the fraction of the
day.}\\
\hline\hline
JD &$ JD+ $&$U$&$\Delta U$&$ JD+ $&$B$&$\Delta B$&$ JD+ $&$V$&$\Delta V$&$ JD+ $&$R_C$&$\Delta R_C$&$ JD+ $&$I_C$&$\Delta I_C$&$ JD+ $&$B-R_C$ &$ JD+ $&$B-V$ &$ JD+ $&$V-R_C$\\
\hline
\endfirsthead
\caption{continued.}\\
\hline\hline
JD &$ JD+ $&$U$&$\Delta U$&$ JD+ $&$B$&$\Delta B$&$ JD+ $&$V$&$\Delta V$&$ JD+ $&$R_C$&$\Delta R_C$&$ JD+ $&$I_C$&$\Delta I_C$&$ JD+ $&$B-R_C$ &$ JD+ $&$B-V$ &$ JD+ $&$V-R_C$\\
\hline
\endhead
\hline
\endfoot

2456783 & --- & --- & --- & .56595 & 11.150 & 0.446 & .56676 & 10.832 & 0.435 & .56758 & 10.533 & 0.315 & .56840 & 10.159 & 0.144 & .56677 & 0.617 & .56636 & 0.318 & .56717 & 0.299\\
2456786 & --- & --- & --- & .54058 & 11.131 & 0.427 & .54195 & 10.793 & 0.396 & .54331 & 10.513 & 0.295 & .54358 & 10.120 & 0.105 & .54195 & 0.618 & .54126 & 0.338 & .54263 & 0.280\\
2456787 & --- & --- & --- & .53645 & 11.146 & 0.442 & .53727 & 10.825 & 0.428 & .53808 & 10.521 & 0.303 & .53909 & 10.160 & 0.145 & .53726 & 0.625 & .53686 & 0.321 & .53767 & 0.304\\
2456799 & .38924 & 10.973 & 0.113 & .39133 & 11.242 & 0.538 & .38942 & 10.828 & 0.431 & .39005 & 10.536 & 0.318 & .39056 & 10.153 & 0.138 & .39069 & 0.706 & .39037 & 0.414 & .38974 & 0.292\\
2456824 & .35969 & 10.965 & 0.105 & .36043 & 11.231 & 0.527 & .36117 & 10.817 & 0.420 & .36180 & 10.526 & 0.308 & .36232 & 10.159 & 0.144 & .36112 & 0.705 & .36080 & 0.414 & .36148 & 0.291\\
2456830 & --- & --- & --- & .51439 & 11.140 & 0.436 & --- & --- & --- & --- & --- & --- & --- & --- & --- & .51658 & 0.598 & --- & --- & --- & ---\\
2456830 & --- & --- & --- & .51714 & 11.162 & 0.458 & .51796 & 10.839 & 0.442 & .51878 & 10.542 & 0.324 & .51959 & 10.162 & 0.147 & .51796 & 0.620 & .51755 & 0.323 & .51837 & 0.297\\
2456835 & .50993 & 10.973 & 0.113 & .51244 & 11.267 & 0.563 & .50688 & 10.853 & 0.456 & .50745 & 10.563 & 0.345 & .50784 & 10.173 & 0.158 & .50995 & 0.704 & .50966 & 0.414 & .50716 & 0.290\\
\end{longtable}

\begin{list}{}{}
\item[Observers:] L. Hambalek, Pribulla, Kundra, Nedoroscik, J. Lopatovsky, E. Kundra, J. Nedoroscik, Z. Garai
\end{list}
\end{landscape}
}

\addtocounter{table}{7}
\longtab[7]{
\begin{landscape}
\begin{longtable}{|lrrrrrrrrrrrrrrrrrr|}
   \caption{\label{TA7} $BV(RI)_{\rm{C}}$ photometry obtained at Furze Hill
(United Kingdom) Observatory during
and near the 2014 eclipse (E = 11).  The 0.35 Schmidt-Cassegrain telescope with
a CCD camera was used.  Comparisons were made against three stars: $BD+55$\,2690,
GSC-3973 2150, $BD+55$\,2691.  The apparent magnitudes are also given with respect
to $BD+55$\,2690 (calculated with the adoption of $BV(RI)_{\rm{C}}$
magnitudes of $BD+55$\,2690 given by Athens Observatory).  The columns
labeled 'JD+' denote the
fraction of the day.}\\

\hline\hline
JD &$ JD+ $&$B$&$\Delta B$&$ JD+ $&$V$&$\Delta V$&$ JD+ $&$R_C$&$\Delta R_C$&$ JD+ $&$I_C$&$\Delta I_C$&$ JD+ $&$B-R_C$ &$ JD+ $&$B-V$ &$ JD+ $&$V-R_C$\\
\hline
\endfirsthead
\caption{continued.}\\
\hline\hline
JD &$ JD+ $&$B$&$\Delta B$&$ JD+ $&$V$&$\Delta V$&$ JD+ $&$R_C$&$\Delta R_C$&$ JD+ $&$I_C$&$\Delta I_C$&$ JD+ $&$B-R_C$ &$ JD+ $&$B-V$ &$ JD+ $&$V-R_C$\\
\hline
\endhead
\hline
\endfoot

2456820 & .47548 & 11.192 & 0.488 & .48224 & 10.824 & 0.427 & .48615 & 10.543 & 0.325 & .49028 & 10.198 & 0.183 & .48081 & 0.649 & .47886 & 0.368 & .48419 & 0.281\\
2456823 & .45086 & 11.196 & 0.492 & .45447 & 10.836 & 0.439 & .45847 & 10.547 & 0.329 & .46238 & 10.196 & 0.181 & .45467 & 0.649 & .45267 & 0.360 & .45647 & 0.289\\
2456826 & .45197 & 11.193 & 0.489 & .45647 & 10.833 & 0.436 & .45979 & 10.543 & 0.325 & .46814 & 10.196 & 0.181 & .45588 & 0.650 & .45422 & 0.360 & .45813 & 0.290\\
2456830 & .44714 & 11.185 & 0.481 & .45155 & 10.826 & 0.429 & .45502 & 10.542 & 0.324 & .45889 & 10.200 & 0.185 & .45108 & 0.643 & .44935 & 0.359 & .45329 & 0.284\\
2456839 & .43823 & 11.197 & 0.493 & .44659 & 10.844 & 0.447 & .45481 & 10.552 & 0.334 & .46400 & 10.208 & 0.193 & .44652 & 0.645 & .44241 & 0.353 & .45070 & 0.292\\
2456845 & .42790 & 11.209 & 0.505 & .43367 & 10.844 & 0.447 & .43572 & 10.561 & 0.343 & .44082 & 10.214 & 0.199 & .43181 & 0.648 & .43079 & 0.365 & .43469 & 0.283\\
2456848 & .42752 & 11.199 & 0.495 & .43177 & 10.838 & 0.441 & .43524 & 10.556 & 0.338 & .44006 & 10.207 & 0.192 & .43138 & 0.643 & .42965 & 0.361 & .43351 & 0.282\\
2456854 & .41838 & 11.206 & 0.502 & .42318 & 10.852 & 0.455 & .42698 & 10.570 & 0.352 & .43163 & 10.214 & 0.199 & .42268 & 0.636 & .42078 & 0.354 & .42508 & 0.282\\
\end{longtable}

\begin{list}{}{}
\item[Observers:] I. Miller
\end{list}
\end{landscape}
}
\addtocounter{table}{8}
\longtab[8]{
\begin{longtable}{|lrrrrrrrr|}
   \caption{\label{TA8} Photometry in standard $BVI$ filters obtained at Szczecin Observatory (Poland)
during and near the 2014 eclipse (E = 11). The 0.2 m Newton telescope with
an Artemis 285 CCD camera was used. Differential magnitudes are given with
respect to $BD+55$\,2690. The columns labeled 'HJD+' denote the fraction part
of the day.}\\
\hline\hline
JD &$ JD+ $&$\Delta B$&$ JD+ $&$\Delta V$&$ JD+ $&$\Delta I$&$ JD+ $&$B-V$\\
\hline
\endfirsthead
\caption{continued.}\\
\hline\hline
JD &$ JD+ $&$\Delta B$&$ JD+ $&$\Delta V$&$ JD+ $&$\Delta I$&$ JD+ $&$B-V$\\
\hline
\endhead
\hline
\endfoot
2456857 & .51180 & 0.495 & --- & --- & --- & --- & --- & ---\\
2456859 & .40070 & 0.498 & .43820 & 0.425 & --- & --- & .41945 & 0.073\\
2456860 & .41810 & 0.492 & .46940 & 0.416 & --- & --- & .44375 & 0.076\\
2456861 & .44440 & 0.503 & .46040 & 0.439 & --- & --- & .45240 & 0.064\\
2456862 & .52360 & 0.502 & .51040 & 0.450 & --- & --- & .51700 & 0.052\\
2456866 & .47010 & 0.500 & .41180 & 0.440 & .51390 & 0.165 & .44095 & 0.060\\
2456868 & .52500 & 0.497 & .51530 & 0.436 & --- & --- & .52015 & 0.061\\
2456870 & .39650 & 0.500 & .40830 & 0.434 & --- & --- & .40240 & 0.066\\
\end{longtable}

\begin{list}{}{}
\item[Observers:] T. Smela
\end{list}
}
\addtocounter{table}{9}
\longtab[9]{
\begin{longtable}{|lrrrrrr|}
   \caption{\label{TA9} $VI_{\rm{C}}$ photometry obtained at West Challow Observatory (United Kingdom) during
and near the 2014 eclipse (E = 11). The 0.356 Schmidt-Cassegrain telescope
with a CCD camera was used. Comparisons were made against three stars:
$BD+55$\,2690, $GSC-3973$\,2150, $BD+55$\,2691. The apparent magnitudes are also given
with respect to $BD+55$\,2690 (calculated with the adoption of $VI_{\rm{C}}$ magnitudes
of $BD+55$\,2690 given by Athens Observatory. Comparison was made against star
$BD+55$\,2690. The columns labeled 'JD+' denote the fraction of the day.}\\
\hline\hline
JD &$ JD+ $&$V$&$\Delta V$&$ JD+ $&$I_C$&$\Delta I_C$\\
\hline
\endfirsthead
\caption{continued.}\\
\hline\hline
JD &$ JD+ $&$V$&$\Delta V$&$ JD+ $&$I_C$&$\Delta I_C$\\
\hline
\endhead
\hline
\endfoot
2456848 & .45307 & 10.819 & 0.422 & .45307 & 10.200 & 0.185\\
2456849 & .46785 & 10.821 & 0.424 & .46785 & 10.211 & 0.196\\
2456852 & .43945 & 10.828 & 0.431 & .43945 & 10.216 & 0.201\\
2456854 & .45390 & 10.828 & 0.431 & .45390 & 10.210 & 0.195\\
2456860 & .43867 & 10.846 & 0.449 & .43867 & 10.228 & 0.213\\
2456862 & .40210 & 10.857 & 0.460 & .40210 & 10.233 & 0.218\\
2456869 & .39980 & 10.849 & 0.452 & .39980 & 10.236 & 0.221\\
2456876 & .37018 & 10.961 & 0.564 & .37018 & 10.327 & 0.312\\
\end{longtable}

\begin{list}{}{}
\item[Observers:] D. Boyd
\end{list}
}
\addtocounter{table}{10}
\longtab[10]{
\begin{longtable}{|lrrr|}
   \caption{\label{TA10} Photometry in standard $V$ filter obtained at 
 Las Pegueras Observatory (Spain) during and near the 2014 eclipse 
(E = 11). The 0.35 Schmidt-Cassegrain telescope with a CCD camera was 
used. Comparisons were made against star $GSC-3973$\,2150. The apparent 
magnitudes are also given with respect to $BD+55$\,2690 (calculated
with the adoption of $V$ magnitudes of $BD+55$\,2690 given in Mikolajewski et al,
2003, IBVS 5412. The columns labeled 'HJD+' denote the fraction of the day.}\\
\hline\hline
$HJD $&$HJD+$&$V$&$V-B$\\
\hline
\endfirsthead
\caption{continued.}\\
\hline\hline
$HJD $&$HJD+$&$V$&$V-B$\\
\hline
\endhead
\hline
\endfoot
2456811 & .53653 & 10.792 & 0.412\\
2456814 & .47751 & 10.829 & 0.449\\
2456817 & .49962 & 10.831 & 0.451\\
2456820 & .48747 & 10.831 & 0.451\\
2456824 & .47660 & 10.831 & 0.451\\
2456828 & .50271 & 10.844 & 0.464\\
2456830 & .52584 & 10.833 & 0.453\\
2456835 & .46567 & 10.846 & 0.466\\
\end{longtable}

\begin{list}{}{}
\item[Observers:] T. A. Heras
\end{list}
}

\addtocounter{table}{11}
\longtab[11]{
\begin{landscape}
\begin{longtable}{|lrrrrrrrrrrrrrrr|}
   \caption{\label{TA11} $UBV(RI)_{\rm{C}}$ photometry obtained at Rohzen (Bulgaria) Observatory during and near
the 2014 eclipse (E = 11). The 0.7 Schmidt telescope with a FLI PL 9000 CCD
camera was used. Comparisons were made against three stars: $BD+55$\,2690,
$GSC-3973$\,2150, $BD+55$\,2691. The apparent magnitudes are also given with respect
to $BD+55$\,2690 (calculated with the adoption of $UBV(RI)_{\rm{C}}$ magnitudes of $BD+55$\,
2690 given by Athens Observatory). Comparison was made against star $BD+55$\,
2690. The columns labeled 'JD+' denote the fraction of the day.}\\
\hline\hline
JD &$ JD+ $&$U$&$\Delta U$&$ JD+ $&$B$&$\Delta B$&$ JD+ $&$V$&$\Delta V$&$ JD+ $&$R_C$&$\Delta R_C$&$ JD+ $&$I_C$&$\Delta I_C$\\
\hline
\endfirsthead
\caption{continued.}\\
\hline\hline
JD &$ JD+ $&$U$&$\Delta U$&$ JD+ $&$B$&$\Delta B$&$ JD+ $&$V$&$\Delta V$&$ JD+ $&$R_C$&$\Delta R_C$&$ JD+ $&$I_C$&$\Delta I_C$\\
\hline
\endhead
\hline
\endfoot

2456838 & .35610 & 10.975 & 0.115 & .35610 & 11.159 & 0.455 & .35610 & 10.819 & 0.422 & .35610 & 10.543 & 0.325 & .35610 & 10.159 & 0.144\\
2456839 & .42961 & 10.983 & 0.123 & .42961 & 11.168 & 0.464 & .42961 & 10.829 & 0.432 & .42961 & 10.549 & 0.331 & .42961 & 10.160 & 0.145\\
2456859 & .48300 & 11.015 & 0.155 & .48300 & 11.203 & 0.499 & .48300 & 10.854 & 0.457 & .48300 & 10.561 & 0.343 & .48300 & 10.174 & 0.159\\
2456860 & .49993 & 11.031 & 0.171 & .49993 & 11.208 & 0.504 & .49993 & 10.874 & 0.477 & .49993 & 10.565 & 0.347 & .49993 & 10.181 & 0.166\\
2456863 & .44882 & 11.033 & 0.173 & .44882 & 11.206 & 0.502 & .44882 & 10.868 & 0.471 & .44882 & 10.587 & 0.369 & .44882 & 10.187 & 0.172\\
2456873 & .37146 & 11.067 & 0.207 & .37146 & 11.243 & 0.539 & .37146 & 10.897 & 0.500 & .37146 & 10.607 & 0.389 & .37146 & 10.225 & 0.210\\
2456874 & .39731 & 11.107 & 0.247 & .39731 & 11.274 & 0.570 & .39731 & 10.935 & 0.538 & .39731 & 10.648 & 0.430 & .39731 & 10.236 & 0.221\\
2456888 & .43818 & 11.512 & 0.652 & .43818 & 11.670 & 0.966 & .43818 & 11.288 & 0.891 & .43818 & 10.976 & 0.758 & .43818 & 10.581 & 0.566\\
\end{longtable}

\begin{list}{}{}
\item[Observers:] E. Semkov
\end{list}
\end{landscape}
}

\addtocounter{table}{12}
\longtab[12]{
\begin{longtable}{|lrrr|}
   \caption{\label{TA12} $V$ photometry obtained at Glyfada Observatory during and near the 2014 eclipse
(E = 11). The 0.28 m Schmidt-Cassegrain telescope with a QHY9 CCD camera was
used. Comparisons were made against star $BD+55$\,2690. The apparent magnitudes
are also given with respect to $BD+55$\,2690 (calculated with the adoption of $V$
magnitudes of $BD+55$\,2690 given in Mikolajewski et al, 2003, IBVS 5412. The
columns labeled 'JD+' denote the fraction of the day.}\\
\hline\hline
JD &$ JD+ $&$V$& $\Delta V$\\
\hline
\endfirsthead
\caption{continued.}\\
\hline\hline
JD &$ JD+ $&$V$& $\Delta V$\\
\hline
\endhead
\hline
\endfoot

2456850 & .32540 & 10.784 & 0.404\\
2456858 & .36738 & 10.810 & 0.43\\
2456859 & .28847 & 10.872 & 0.492\\
2456873 & .33900 & 10.899 & 0.519\\
2456874 & .34757 & 10.927 & 0.547\\
2456877 & .43292 & 10.952 & 0.572\\
2456881 & .29096 & 11.034 & 0.654\\
2456882 & .28657 & 11.051 & 0.671\\
\end{longtable}

\begin{list}{}{}
\item[Observers:] M. Kardasis
\end{list}
}
\addtocounter{table}{13}
\longtab[13]{
\begin{longtable}{|lrrrrrr|}
   \caption{\label{TA13} Photometry obtained at PTMA Szczecin Observatory (Poland) in standard $BV$
filters during and near the 2014 eclipse (E = 11). The 0.2 m Newton telescope
with an Artemis 285 CCD camera was used. Differential magnitudes are given
with respect to $BD+55$\,2690. The columns labeled 'HJD+' denote the fraction
of the day.}\\
\hline\hline
JD &$ JD+ $&$\Delta B$&$ JD+ $&$\Delta V$&$ JD+ $&$B-V$\\
\hline
\endfirsthead
\caption{continued.}\\
\hline\hline
JD &$ JD+ $&$\Delta B$&$ JD+ $&$\Delta V$&$ JD+ $&$B-V$\\
\hline
\endhead
\hline
\endfoot
2456859 & .39860 & 0.506 & --- & --- & --- & --- \\
2456866 & .47710 & 0.486 & .41320 & 0.424 & .44515 & 0.062\\
2456868 & --- & --- & .52080 & 0.442 & --- & --- \\
2456872 & .50070 & 0.532 & .49030 & 0.468 & .49550 & 0.064\\
2456875 & .56740 & 0.635 & --- & --- & --- & --- \\
2456876 & --- & --- & .46810 & 0.550 & --- & --- \\
2456877 & .57780 & 0.639 & --- & --- & --- & --- \\
2456878 & --- & --- & .45280 & 0.555 & --- & --- \\
\end{longtable}

\begin{list}{}{}
\item[Observers:] M. Biskupski
\end{list}
}
\addtocounter{table}{14}
\longtab[14]{
\begin{longtable}{|lrrrrrrrrr|}
   \caption{\label{TA14} $BVR_{\rm{C}}$ photometry obtained at Spain Observatory during and near the 2014
eclipse (E = 11). The 0.431 Corrected Dall-Kirkham telescope with a
SBIG STL-11000M CCD camera was used. Comparisons were made against three
stars: $BD+55$\,2690, $GSC-3973$\,2150, $BD+55$\,2691. The apparent magnitudes are also
given with respect to $BD+55$\,2690 (calculated with the adoption of $BVR_{\rm{C}}$
magnitudes of $BD+55$\,2690 given by Athens Observatory). Comparison was made
against star $BD+55$\,2690. The columns labeled 'JD+' denote the fraction of the
day.}\\
\hline\hline
JD &$ JD+ $&$B$&$\Delta B$&$ JD+ $&$V$&$\Delta V$&$ JD+ $&$R_C$& $\Delta R_C$\\
\hline
\endfirsthead
\caption{continued.}\\
\hline\hline
JD &$ JD+ $&$B$&$\Delta B$&$ JD+ $&$V$&$\Delta V$&$ JD+ $&$R_C$& $\Delta R_C$\\
\hline
\endhead
\hline
\endfoot
2456850 & .87207 & 11.185 & 0.481 & ---    & ---    & ---   & ---    & ---    & ---\\
2456852 & .53206 & 11.180 & 0.476 & .52920 & 10.841 & 0.444 & .53495 & 10.582 & 0.364\\
2456855 & --- & --- & --- & .92490 & 10.842 & 0.445 & ---   & ---    & --- \\
2456857 & --- & --- & --- & .39680 & 10.855 & 0.458 & --- & --- & --- \\
2456859 & --- & --- & --- & .53700 & 10.855 & 0.458 & --- & --- & --- \\
2456859 & --- & --- & --- & .57560 & 10.849 & 0.452 & --- & --- & --- \\
2456860 & --- & --- & --- & .57660 & 10.854 & 0.457 & --- & --- & --- \\
2456862 & --- & --- & --- & .58230 & 10.859 & 0.462 & --- & --- & --- \\
\end{longtable}

\begin{list}{}{}
\item[Observers:] R. Kneip
\end{list}
}
\addtocounter{table}{15}
\longtab[15]{
\begin{longtable}{|lrrrrrrrrrrrrr|}
   \caption{\label{TA15} $BVR_{\rm{C}}$ photometry obtained at Guadarrama (Spain) Observatory during and near
the 2014 eclipse (E = 11). The 0.25 Schmidt-Cassegrain telescope with a SBIG
ST8XME CCD camera was used. Comparisons were made against three stars: $BD+55$\,2690, $GSC-3973$\,2150, $BD+55$\,2691. The apparent magnitudes are also given with
respect to $BD+55$\,2690 (calculated with the adoption of $BV(RI)_{\rm{C}}$ magnitudes of
$BD+55$\,2690 given by Athens Observatory). The columns labeled 'JD+' denote the
fraction of the day.}\\
\hline\hline
JD &$ JD+ $&$B$&$\Delta B$&$ JD+ $&$V$&$\Delta V$&$ JD+ $&$R_C$&$\Delta R_C$&$ JD+ $&$B-V$ &$ JD+ $&$V-R_C$\\
\hline
\endfirsthead
\caption{continued.}\\
\hline\hline
JD &$ JD+ $&$B$&$\Delta B$&$ JD+ $&$V$&$\Delta V$&$ JD+ $&$R_C$&$\Delta R_C$&$ JD+ $&$B-V$ &$ JD+ $&$V-R_C$\\
\hline
\endhead
\hline
\endfoot

2456824 & --- & --- & --- & .51935 & 10.828 & 0.431 & .52410 & 10.578 & 0.360 & --- & --- & --- & --- \\
2456844 & --- & --- & --- & .50280 & 10.825 & 0.428 & .50383 & 10.557 & 0.339 & --- & --- & .50331 & 0.268\\
2456847 & .48485 & 11.149 & 0.445 & .47808 & 10.831 & 0.434 & --- & --- & --- & .48146 & 0.318 & .52172 & 0.250\\
2456853 & --- & --- & --- & .43723 & 10.840 & 0.443 & .43862 & 10.545 & 0.327 & --- & --- & .43792 & 0.295\\
2456862 & --- & --- & --- & .45141 & 10.851 & 0.454 & .45359 & 10.557 & 0.339 & --- & --- & .45250 & 0.294\\
2456868 & --- & --- & --- & .44130 & 10.864 & 0.467 & .44329 & 10.585 & 0.367 & --- & --- & .44229 & 0.279\\
2456874 & --- & --- & --- & .42506 & 10.916 & 0.519 & .42350 & 10.634 & 0.416 & --- & --- & .42428 & 0.282\\
2456880 & --- & --- & --- & .42607 & 11.038 & 0.641 & .43028 & 10.763 & 0.545 & --- & --- & .42818 & 0.275\\
\end{longtable}

\begin{list}{}{}
\item[Observers:] D. Rodriguez
\end{list}
}

\addtocounter{table}{16}
\longtab[16]{
\begin{longtable}{|lrrr|}
   \caption{\label{TA16} $V$ photometry obtained at Zillebeke (Belgium) during and near the 2014 eclipse (E = 11).
The 0.68 m Newton telescope with an SBIG STL-6303E CCD camera was used.
Comparisons were made against three stars: $BD+55$\,2690, $GSC-3973$\,2150,
$BD+55$\,2691. The apparent magnitudes are also given with respect to $BD+55$\,2690
(calculated with the adoption of $V$ magnitudes of $BD+55$\,2690 given in
Mikolajewski et al, 2003, IBVS 5412. The columns labeled 'JD+' denote the
fraction of the day.}\\
\hline\hline
JD &$ JD+ $&$V$& $\Delta V$\\
\hline
\endfirsthead
\caption{continued.}\\
\hline\hline
JD &$ JD+ $&$V$& $\Delta V$\\
\hline
\endhead
\hline
\endfoot
2456822 & .41101 & 10.810 & 0.430\\
2456829 & .39743 & 10.820 & 0.440\\
2456832 & .41820 & 10.807 & 0.427\\
2456838 & .44653 & 10.815 & 0.435\\
2456842 & .44393 & 10.823 & 0.443\\
2456845 & .44125 & 10.824 & 0.444\\
2456853 & .40167 & 10.840 & 0.460\\
2456856 & .50551 & 10.840 & 0.460\\
2456861 & .38245 & 10.845 & 0.465\\
2456868 & .37139 & 10.844 & 0.464\\
2456871 & .36254 & 10.853 & 0.473\\
2456873 & .38770 & 10.888 & 0.508\\
2456876 & .40966 & 10.961 & 0.581\\
2456880 & .36377 & 11.018 & 0.638\\
2456885 & .34749 & 11.093 & 0.713\\
2456889 & .39466 & 11.357 & 0.977\\
2456892 & .34973 & 11.464 & 1.084\\
2456896 & .62099 & 11.575 & 1.195\\
2456897 & .33568 & 11.321 & 0.941\\
2456898 & .32146 & 11.363 & 0.983\\
2456901 & .33859 & 11.169 & 0.789\\
2456903 & .32816 & 11.096 & 0.716\\
2456907 & .35213 & 10.889 & 0.509\\
2456909 & .31654 & 10.847 & 0.467\\
2456912 & .31542 & 10.834 & 0.454\\
\end{longtable}

\begin{list}{}{}
\item[Observers:] F. Dubois
\end{list}
}
\addtocounter{table}{17}
\longtab[17]{
\begin{longtable}{|lrrrrrrrr|}
   \caption{\label{TA17} Photometry obtained at Rozhen Observatory
(Bulgaria) in standard $BV(RI)_{\rm{C}}$ filters during and near the 2014
eclipse (E = 11).  The 0.6 m Cassegrain telescope with a FLI PL 9000 CCD
camera was used.  Differential magnitudes are given with respect to
$BD+55$\,2690.  The columns labeled 'JD+' denote the fraction of the
day.}\\
\hline\hline
JD &$ JD+ $&$\Delta B$&$ JD+ $&$\Delta V$&$ JD+ $&$\Delta R_C$&$ JD+ $&$\Delta I_C$\\
\hline
\endfirsthead
\caption{continued.}\\
\hline\hline
JD &$ JD+ $&$\Delta B$&$ JD+ $&$\Delta V$&$ JD+ $&$\Delta R_C$&$ JD+ $&$\Delta I_C$\\
\hline
\endhead
\hline
\endfoot
2456865 & .54979 & 0.497 & .55260 & 0.460 & .54670 & 0.348 & .55906 & 0.159\\
2456866 & .55503 & 0.489 & .55722 & 0.451 & .55885 & 0.346 & .56049 & 0.156\\
2456890 & .58708 & 1.117 & .59086 & 1.047 & .58452 & 0.894 & --- & ---\\
2456890 & .59653 & 1.128 & .59984 & 1.049 & .60226 & 0.893 & --- & ---\\
2456891 & .57730 & 1.145 & .57992 & 1.072 & .57503 & 0.921 & --- & ---\\
2456891 & .58583 & 1.147 & .58900 & 1.067 & .59110 & 0.912 & --- & ---\\
2456894 & .43321 & 1.219 & .43608 & 1.164 & .43828 & 0.999 & --- & ---\\
\end{longtable}

\begin{list}{}{}
\item[Observers:] G. Apostolovska
\end{list}
}

\addtocounter{table}{18}
\longtab[18]{
\tiny{
\begin{landscape}
\begin{longtable}{|lrrrrrrrrrrrrrrrrrr|}
   \caption{\label{TA18} $BV(RI)_{\rm{C}}$ photometry obtained at Tenerife
(Spain) Observatory during and near the 2014 eclipse (E = 11).  The 0.356
Schmidt-Cassegrain telescope with a FLI MicroLine CCD camera was used.
Comparisons were made against three stars: $BD+55$\,2690, $GSC-3973$\,2150,
$BD+55$\,2691.  The apparent magnitudes are also given with respect to
$BD+55$\,2690 (calculated with the adoption of $BV(RI)_{\rm{C}}$ magnitudes
of $BD+55$\,2690 given by Athens Observatory). Comparison was made against
star $BD+55$\,2690.  The columns labeled 'JD+' denote the fraction of the
day.}\\
\hline\hline
JD &$ JD+ $&$B$&$\Delta B$&$ JD+ $&$V$&$\Delta V$&$ JD+ $&$R_C$&$\Delta R_C$&$ JD+ $&$I_C$&$\Delta I_C$&$ JD+ $&$B-R_C$ &$ JD+ $&$B-V$ &$ JD+ $&$V-R_C$\\
\hline
\endfirsthead
\caption{continued.}\\
\hline\hline
JD &$ JD+ $&$B$&$\Delta B$&$ JD+ $&$V$&$\Delta V$&$ JD+ $&$R_C$&$\Delta R_C$&$ JD+ $&$I_C$&$\Delta I_C$&$ JD+ $&$B-R_C$ &$ JD+ $&$B-V$ &$ JD+ $&$V-R_C$\\
\hline
\endhead
\hline
\endfoot
2456858 & ---    & ---    & ---   & .65831 & 10.850 & 0.453 & .65933 & 10.540 & 0.322 & ---    & ---    & ---   & ---    & ---   & ---    & ---   & ---    & ---  \\
2456864 & .65036 & 11.220 & 0.516 & ---    & ---    & ---   & ---    & ---    & ---   & ---    & ---    & ---   & ---    & ---   & ---    & ---   & ---    & ---  \\
2456867 & ---    & ---    & ---   & .63928 & 10.860 & 0.463 & ---    & ---    & ---   & ---    & ---    & ---   & ---    & ---   & ---    & ---   & ---    & ---  \\
2456874 & ---    & ---    & ---   & .66275 & 10.940 & 0.543 & ---    & ---    & ---   & ---    & ---    & ---   & ---    & ---   & ---    & ---   & ---    & ---  \\
2456880 & ---    & ---    & ---   & .63422 & 11.060 & 0.663 & .63100 & 10.730 & 0.512 & ---    & ---    & ---   & ---    & ---   & ---    & ---   & .63261 & 0.330\\
2456883 & .64933 & 11.480 & 0.776 & ---    & ---    & ---   & ---    & ---    & ---   & .58116 & 10.420 & 0.405 & ---    & ---   & ---    & ---   & ---    & ---  \\
2456891 & ---    & ---    & ---   & .56484 & 11.480 & 1.083 & .56608 & 11.140 & 0.922 & .56351 & 10.730 & 0.715 & ---    & ---   & ---    & ---   & .56546 & 0.340\\
2456897 & ---    & ---    & ---   & ---    & ---    & ---   & ---    & ---    & ---   & .61608 & 10.670 & 0.655 & ---    & ---   & ---    & ---   & ---    & ---  \\
\end{longtable}

\begin{list}{}{}
\item[Observers:] A. Capetillo Blanco
\end{list}
\end{landscape}
}
}

\addtocounter{table}{19}
\longtab[19]{
\begin{landscape}
\begin{longtable}{|lrrrrrrrrrrrrrr|}
   \caption{\label{TA19} Photometry in standard $UBVR$ filters obtained at Jan Kochanowski Observatory (Poland)
during and near the 2014 eclipse (E = 11). The 0.35 m Schmidt-Cassegrain
telescope with a SBIG ST-7XE CCD camera was used.  Differential magnitudes
are given with respect to $BD+55$\,2690. The columns labeled 'JD+' denote the
fraction part of the day.}\\
\hline\hline
JD &$ JD+ $&$\Delta U$&$HJD+$&$\Delta B$&$ JD+ $&$\Delta V$&$ JD+ $&$\Delta R$&$ JD+ $&$B-R_C$ &$ JD+ $&$B-V$ &$ JD+ $&$V-R_C$\\
\hline
\endfirsthead
\caption{continued.}\\
\hline\hline
JD &$ JD+ $&$\Delta U$&$HJD+$&$\Delta B$&$ JD+ $&$\Delta V$&$ JD+ $&$\Delta R$&$ JD+ $&$B-R_C$ &$ JD+ $&$B-V$ &$ JD+ $&$V-R_C$\\
\hline
\endhead
\hline
\endfoot
2456799 & .36752 & 0.157 & .37804 & 0.470 & .38497 & 0.421 & .38925 & 0.316 & .38364 & 0.154 & .38150 & 0.049 & .38711 & 0.105\\
2456820 & .34564 & 0.148 & .35475 & 0.429 & .33470 & 0.413 & .33892 & 0.298 & .34683 & 0.131 & .34472 & 0.016 & .33681 & 0.115\\
2456892 & .33113 & 0.892 & .34005 & 1.189 & .31967 & 1.055 & .32369 & 0.907 & .33187 & 0.282 & .32986 & 0.134 & .32168 & 0.148\\
2456895 & .31833 & 0.954 & .31833 & 1.037 & .31833 & 1.098 & .28860 & 0.952 & .30347 & 0.085 & .31833 &-0.061 & .30347 & 0.146\\
2456905 & .29905 & 0.386 & .31709 & 0.660 & .28324 & 0.596 & .28740 & 0.484 & .30224 & 0.176 & .30016 & 0.064 & .28532 & 0.112\\
\end{longtable}

\begin{list}{}{}
\item[Observers:] P. Kankiewicz
\end{list}
\end{landscape}
}
\addtocounter{table}{20}
\longtab[20]{
\begin{longtable}{|lrrrrrrrrrrr|}
   \caption{\label{TA20} $BVI_{\rm{C}}$ photometry obtained at Magnago (Italy) Observatory during and near the
2014 eclipse (E = 11). The 0.25 Schmidt-Cassegrain telescope with a SBIG ST9
CCD camera was used. Comparisons were made against star $BD+55$\,2690. The
apparent magnitudes are also given with respect to $BD+55$\,2690 (calculated
with the adoption of $BVI_{\rm{C}}$ magnitudes of $BD+55$\,2690 given by Athens
Observatory). Comparison was made against star $BD+55$\,2690. The columns
labeled 'JD+' denote the fraction of the day.}\\
\hline\hline
JD &$ JD+ $&$B$&$\Delta B$&$ JD+ $&$V$&$\Delta V$&$ JD+ $&$I_C$&$\Delta I_C$&$ JD+ $&$B-V$\\
\hline
\endfirsthead
\caption{continued.}\\
\hline\hline
JD &$ JD+ $&$B$&$\Delta B$&$ JD+ $&$V$&$\Delta V$&$ JD+ $&$I_C$&$\Delta I_C$&$ JD+ $&$B-V$\\
\hline
\endhead
\hline
\endfoot
2456801 & .59070 & 11.175 & 0.471 & .59150 & 10.819 & 0.422 & .59200 & 10.147 & 0.132 & .59110 & 0.356\\
2456842 & .47750 & 11.170 & 0.466 & .47860 & 10.810 & 0.413 & .47910 & 10.167 & 0.152 & .47805 & 0.360\\
2456854 & .48830 & 11.173 & 0.469 & .48940 & 10.847 & 0.450 & .48990 & 10.199 & 0.184 & .48885 & 0.326\\
2456903 & .45240 & 11.462 & 0.758 & .45360 & 11.089 & 0.692 & .45410 & 10.450 & 0.435 & .45300 & 0.373\\
2456933 & .46070 & 11.156 & 0.452 & .46180 & 10.845 & 0.448 & .46230 & 10.186 & 0.171 & .46125 & 0.311\\
\end{longtable}

\begin{list}{}{}
\item[Observers:] M. Martigoni
\end{list}
}
\longtab[21]{
\begin{longtable}{|lrrrrrrrrrrrrrrr|}
   \caption{\label{TA21} $BVR_{\rm{B}}$ photometry in Bessell obtained at Horten VGS (Norway) Observatory during and near
the 2014 eclipse (E = 11). The 0.46 Ritchey-Chretien telescope with a SBIG
STL-1301E CCD camera was used. Comparisons were made against three stars:
$BD+55$\,2690, $GSC-3973$\,2150, $BD+55$\,2691. The apparent magnitudes are also given
with respect to $BD+55$\,2690 (calculated with the adoption of $BV(RI)_{\rm{C}}$
magnitudes of $BD+55$\,2690 given by Athens Observatory). The columns labeled
'JD+' denote the fraction of the day.}\\
\hline\hline
JD &$ JD+ $&$B$&$\Delta B$&$ JD+ $&$V$&$\Delta V$&$ JD+ $&$R_C$&$\Delta R_C$&$ JD+ $&$B-R_C$ &$ JD+ $&$B-V$ &$ JD+ $&$V-R_C$\\
\hline
\endfirsthead
\caption{continued.}\\
\hline\hline
JD &$ JD+ $&$B$&$\Delta B$&$ JD+ $&$V$&$\Delta V$&$ JD+ $&$R_C$&$\Delta R_C$&$ JD+ $&$B-R_C$ &$ JD+ $&$B-V$ &$ JD+ $&$V-R_C$\\
\hline
\endhead
\hline
\endfoot
2456895 & .33888 & 11.842 & 1.138 & --- & --- & --- & --- & --- & --- & --- & --- & --- & --- & --- & --- \\
2456895 & --- & --- & --- & .33986 & 11.460 & 1.063 & .33783 & 11.135 & 0.917 & .33836 & 0.707 & .33937 & 0.382 & .33885 & 0.325\\
2456896 & --- & --- & --- & .35777 & 11.460 & 1.063 & .35714 & 11.124 & 0.906 & --- & --- & --- & --- & --- & --- \\
2456901 & .41431 & 11.546 & 0.842 & .41207 & 11.184 & 0.787 & .41375 & 10.890 & 0.672 & .41403 & 0.656 & .41319 & 0.362 & .35746 & 0.336\\
2456906 & .32484 & 11.291 & 0.587 & .32369 & 10.990 & 0.593 & .32419 & 10.682 & 0.464 & --- & --- & .32426 & 0.301 & .41291 & 0.294\\
2456906 & --- & --- & --- & --- & --- & --- & --- & --- & --- & --- & --- & --- & --- & .32394 & 0.308\\

\end{longtable}

\begin{list}{}{}
\item[Observers:] J. Kare Trandem Qvam
\end{list}
}
\addtocounter{table}{22}
\longtab[22]{
%

\end{landscape}
}
}
\end{appendix}

\end{document}